\documentclass[universe,article,accept,moreauthors,pdftex]{Definitions/mdpi} 
\usepackage[finalnew]{trackchanges}
\firstpage{1} 
\pubvolume{9}
\issuenum{4}
\articlenumber{169}
\pubyear{2023}
\copyrightyear{2023}
\externaleditor{Academic Editor: Sergei D. Odintsov}

\datereceived{24 February 2023 } 
\daterevised{27 March 2023 } 
\dateaccepted{28 March 2023} 
\datepublished{30 March 2023} 
\hreflink{https://
doi.org/10.3390/universe9040169} 

\newcommand{\ord}[1]{\mathcal{O} \left(#1\right)}
\newcommand{\dtau}[1]{\frac{\partial #1}{\partial \tau}}
\newcommand{\dT}[1]{\frac{\partial #1}{\partial T}}
\newcommand{\dhij}[2]{\frac{\partial #1}{\partial h_{#2}}}
\newcommand{\dphi}[1]{\frac{\partial #1}{\partial \phi}}
\newcommand{\dphisq}[1]{\frac{\partial^2 #1}{\partial \phi^2}}
\newcommand{\Hubble}[0]{\mathrm{H}}

\Title{Study of the Inflationary Spectrum in the Presence of Quantum Gravity Corrections}

\TitleCitation{Study of the Inflationary Spectrum in the Presence of Quantum Gravity Corrections}

\Author{{Giulia Maniccia} 
 $^{1,2,}$*\orcidG{}, Giovanni Montani $^{3,1}$ and Leonardo Torcellini $^{1}$}

\AuthorNames{Giulia Maniccia, Giovanni Montani and Leonardo Torcellini}
\AuthorCitation{{Maniccia, G.;} 
 Montani, G.; Torcellini, L.}

\address{$^{1}$ \quad Physics Department, “La Sapienza” University of Rome, P.le A. Moro 5, 00185 Roma, Italy; giovanni.montani@enea.it (G.M.); torcellini.1756394@studenti.uniroma1.it (L.T.)\\
$^{2}$ \quad INFN Section of Rome, “La Sapienza” University of Rome, P.le A. Moro 5, 00185 Roma, Italy\\
$^{3}$ \quad FNS Department, ENEA, C.R. Frascati, Via E. Fermi 45, 00044 Frascati, Italy
}

\corres{Correspondence: giulia.maniccia@roma1.infn.it} 

\abstract{
After a brief review of the different approaches to predicting the possible quantum gravity corrections to quantum field theory, we discuss in some detail the formulation based on a \remove{fixed} 
\change{Gaussian reference frame}{Gaussian reference frame fixing}
. Then, we utilize this scenario in the determination of the inflationary spectrum of primordial perturbations.
We consider the quantization of an inhomogeneous, free, massless scalar field in a quasi-classical isotropic Universe by developing a WKB expansion of the dynamics of the next order in the Planckian parameter, with respect to the one at which standard QFT emerges. The quantum gravity corrections to the scale-invariant spectrum are discussed in a specific primordial cosmological setting and then in a general minisuperspace formalism, showing that there is no mode-dependent effect, and thus the scale invariant inflationary spectrum is preserved. 
This result is discussed in connection to the absence of a matter backreaction on the gravitational background in the considered paradigm.
}


\begin{document}

\section{Introduction}\label{sec:intro}

The quantization of the gravitational field is one of the most challenging open questions in modern theoretical physics~\cite{bib:thiemann-book,bib:cianfrani-canonicalQuantumGravity}. In~particular, the~implementation of the canonical method to quantize geometrodynamics~\cite{bib:dewitt1-1967,bib:dewitt2-1967,bib:dewitt3-1967,bib:kuchar-1981} has encountered a number of non-trivial difficulties, among~which is the the so-called "problem of time" \cite{bib:isham-book-1993,bib:wald-1993,bib:rovelli-1991-time,bib:kuchar2011-review}, which~\remove{which} is associated the possibility of constructing a viable Hilbert space for the gravitational states. 
In this respect, significant progress has been achieved by adopting the Ashtekar formulation~\cite{bib:ashtekar-1986,bib:rovelli-book-QG}, which allowed the authors, via the introduction of non-local variables, to~achieve the set-up of a kinematical Hilbert space and to demonstrate the discrete nature of the geometrical operator spectrum~\cite{bib:rovelli-smolin-1995}. 
Nonetheless, some important shortcomings also affect this proposal, dubbed loop quantum gravity, such as the impossibility of coherently implementing the system dynamics and the difficulty of defining a proper classical limit~\cite{bib:nicolai-review-2005}. 

On the contrary, many important achievements have been reached in treating quantum field theory (QFT) on curved spacetime~\cite{bib:birrel-davies,bib:book-wald-QFT-curved-space,bib:wald-review-1995}, such as the derivation of the Unruh and Hawking effects~\cite{bib:crispino-review-unruh-2008,bib:hawking-1975}.
Despite the absence of an experimental confirmation, this new physics has been derived by various different approaches and appears as a well-grounded paradigm, although~there is not yet a general unique formulation in terms of a generic gravitational~field. 

Among the two research fields mentioned above, an~intermediate point of view should receive more attention, i.e.,~the description of possible quantum gravity corrections to QFT on curved spacetime. In~fact, there are physical settings, both in early cosmology and for the gravitational collapse, in~which the background metric cannot be regarded as a purely classical dynamics, but~simply an essentially classical dynamics affected by quantum fluctuations of the geometry.
This line of research has been till now developed by a relatively limited number of studies~\cite{bib:vilenkin-1989,bib:kiefer-1991,bib:barvinsky-1993,bib:vilenkin-1994-rassegna,bib:okhuwa-1995,bib:bertoni-venturi-1996,bib:brizuela-kiefer-2016-desitter,bib:brizuela-kiefer-2016-slow-roll,bib:venturi-2017,bib:kiefer-2018,bib:venturi-2020,bib:rotondo-2020,bib:gielen-2021,bib:montani-digioia-maniccia-2021,bib:maniccia-montani-2021,bib:maniccia-montani-2022,bib:maniccia-deangelis-montani-review-2022}, in~which the first problem to be addressed is the emergence of a time variable for QFT from the gravity--matter Wheeler--DeWitt~equation. 

In a pioneering approach~\cite{bib:rubakov-lapchinsky-1979}, the~so-called "Tomonaga time" was introduced to reconstruct QFT on curved spacetime from canonical quantum gravity. 
However, the~most interesting proposal comes from the well-known analysis~\cite{bib:vilenkin-1989}, treating the separation of the total Hamiltonian as a quasi-classical component and a "small" quantum subsystem (see~\cite{bib:agostini-cianfrani-montani-2017} for a physical characterization of the word "small" in this context). 
A number of interesting implementations of this idea in the cosmological arena can be found in~\cite{bib:banks-1985,bib:vilenkin-1994-rassegna,bib:vilenkin-2002,bib:battisti-belvedere-montani-2009,bib:kiefer-2013-review,bib:brizuela-kiefer-2016-desitter,bib:brizuela-kiefer-2016-slow-roll,bib:moriconi-montani-2017,bib:venturi-2017,bib:montani-marchi-moriconi-2018,bib:damour-vilenkin-2019,bib:kiefer-2019,bib:montani-chiovoloni-cascioli-2020,bib:deangelis-2020,bib:robles-perez-2021,bib:maniccia-montani-2021,bib:peter-kiefer-2022,bib:maniccia-deangelis-montani-review-2022}, {whose common trait is the description of a} "small" quantum subsystem{,} often identified in the Universe anisotropic degrees of freedom in contrast to a quasi-classical isotropic background. Other approaches aimed at obtaining QFT on curved spacetime\change{. Effective theory limits}{as an effective theory limit} from quantum gravity can be found in {Refs.} \cite{bib:sahlmann-thiemann-2006-1,bib:sahlmann-thiemann-2006-2,bib:ashtekar-leewandowski-2009,bib:lewandowski-2012,bib:bojowald-2018}. 

In~\cite{bib:kiefer-1991}, the~original proposal of~\cite{bib:vilenkin-1989} {was} specialized to the problem of quantum gravity corrections to QFT, and overall extended to the next order of approximation, where such a feature affects the QFT functional Schr\"odinger equation. 
This study  had the merit of outlining the emergence of a non-unitary theory, as the QFT Hamiltonian is amended by quantum gravity contributions, at~the first order of expansion in the inverse of a Planckian parameter. This question of the non-unitarity {was} discussed in {later works} \cite{bib:kiefer-2018,bib:chataignier-kramer-2021,bib:gielen-2022}, and {the Born--Oppenheimer character of the adopted scheme was emphasized} in~\cite{bib:bertoni-venturi-1996,bib:venturi-2017,bib:venturi-2020}. 
First in~\cite{bib:montani-digioia-maniccia-2021} and subsequently in~\cite{bib:maniccia-montani-2022}, it was argued that{,} to properly deal with the non-unitarity puzzle, it is necessary to define the physical clock from a different dynamical setting. 
In fact, in~\cite{bib:vilenkin-1989,bib:kiefer-1991} {(see} \cite{bib:montani-digioia-maniccia-2021} {for a detailed comparison of the two works)}, the~time dependence of the QFT wave functional is essentially recovered from the corresponding dependence on the label time of the quasi-classical metric variables. 
It is just in this feature that the non-unitarity naturally manifests itself in the perturbation scheme (for a review of the entire line of research, including some minisuperspace applications see~\cite{bib:maniccia-deangelis-montani-review-2022}). 
In \cite{bib:montani-digioia-maniccia-2021}, {the presence of the so-called kinematical action was postulated in the quantum gravity--matter dynamics} \cite{bib:kuchar-1981,bib:montani-2002}, and in~\cite{bib:maniccia-montani-2022} the whole problem {was} restated in the framework of~\cite{bib:kuchar-torre-1991}, i.e.,~by fixing a Gaussian reference frame  which is "materialized" in the dynamics (see also~\cite{bib:mercuri-montani-2004-framefixing}). 
The idea is that, in~the considered Born--Oppenheimer scenario, the~non-physical nature of the emerging fluid (i.e., its violation of the so-called "energy conditions") does not take place, as~a result of the perturbative~expansion. 

In the present study, we re-analyze this idea and then apply it to the natural arena of predicting possible quantum gravity corrections to the inflationary spectrum. 
The origin of the primordial perturbation spectrum is identified in the quantum fluctuations of the inflaton field during the slow-rolling phase \cite{bib:montani-primordialcosmology,bib:weinberg,bib:kolb-turner,bib:riotto-2017}, here approximated by an exact de Sitter regime~\cite{bib:brizuela-kiefer-2016-desitter}. 
More specifically, we consider a Robertson--Walker quasi-classical background, described via the conformal time, and we study the resulting spectrum of a free massless scalar field living on a de Sitter phase of the Universe, dominated by the vacuum energy of the transition phase, here represented by a cosmological constant term. 
By a Fourier decomposition of the scalar mode, we are able to deal with a set of minisuperspace models, one for each value of the wavenumber. The~Schr\"odinger equation for QFT we consider here is the one obtained in~\cite{bib:maniccia-montani-2022}---see also~\cite{bib:maniccia-montani-2021}---and~the aim of the present study is to evaluate how such a correction can affect the spectrum of the inflaton~field. 

The main result of the considered cosmological scenario is showing how the quantum gravity corrections manifest by a simple phase factor in front of the standard QFT solution on a Robertson--Walker metric, {de facto} corresponding to the solution of a time-dependent harmonic oscillator. As~a natural consequence, the~effect of the quantum gravity corrections on the spectrum vanishes at the considered order of expansion. 
The explanation for such a surprising issue is then discussed in a more general minisuperspace scheme, without~a specific reference to the dynamical setting. We remark that the obtained result depends on the possibility of always factorizing the quantum gravity correction to the Universe wave function with respect to the standard QFT state on the considered cosmological background. 
The physical motivation for such a decoupling of the wave function is finally identified in the absence, up~to first order of approximation, of~a backreaction of the quantum matter on the quasi-classical~background. 

Thus, our study has the main merit of clarifying how, in~the framework proposed in~\cite{bib:maniccia-montani-2022}, the~phenomenology (here identified in the primordial spectrum) is not affected by quantum gravity modifications, at~least up to the considered expansion order, and how this perspective has to be sensitive to the existence of appreciable feedback of the quantum matter dynamics on its background~variables. 

The paper is structured as follows. In~Section~\ref{sec:fluid}, we discuss the implementation of the Gaussian frame procedure to define a time parameter, reviewing the original formulation in Section~\ref{ssec:KT} and the WKB implementation in Section~\ref{ssec:fluidWKB}. In~Section~\ref{sec:powerspectrum}, we apply the considered formalism to calculate the possible corrections to the inflationary spectrum by introducing a Fourier decomposition of the inflaton and determining the vacuum expectation values. In~Section~\ref{sec:applic-gen}, we discuss the possible physical motivations for dealing with the standard (not modified by quantum gravity) inflationary spectrum discussed in Section~\ref{sec:powerspectrum}. The~concluding remarks are presented in Section~\ref{sec:conclusions}.


\section{Reference Frame Fixing and~Reparametrization}\label{sec:fluid}

We discuss here the reparametrization procedure illustrated in~\cite{bib:kuchar-torre-1991} that allowed us to define a physical clock for the quantum gravity system. The~original paradigm was then applied to the case of gravity and matter via a Wentzel--Kramer--Brillouin (WKB) expansion in a Planckian parameter, as~shown in~\cite{bib:maniccia-montani-2022}.

\subsection{Kuchar--Torre Gaussian Frame~Proposal}\label{ssec:KT}
A proposal to recover a physical clock for the quantum gravity system was discussed in~\cite{bib:kuchar-torre-1991}, based on a Gaussian reference frame implementation. The~formalism there introduced allows one to fix the Gaussian frame in a quantum field theory, by~a reparametrization procedure that preserves the system's invariance under the coordinate~\change{option}{choice}.

For this purpose, the~following term is adjoined to the action of the system:
\begin{linenomath}
\begin{equation}\label{eq:S-fluid-param}
    S^f = \int d^4x \left[ \frac{\sqrt{-g}}{2}\,  \left(g^{\alpha \beta} \partial_{\alpha}T(x) \, \partial_{\beta}T(x) -1 \right)\mathcal{F} +\sqrt{-g}\, \left( g^{\alpha \beta} \partial_{\alpha}T(x)\, \partial_{\beta}X^i(x) \right) \mathcal{F}_i  \right]\, 
\end{equation}
\end{linenomath}
where $\mathcal{F}, \mathcal{F}_i$ are Lagrange multipliers. Here, {$T, X^i$ are the Gaussian coordinates} associated with the metric {$\gamma_{\mu\nu}$ satisfying $\gamma^{00}=1, \gamma^{0i}=0$ (the implemented signature is $(+,-,-,-)$ for coherence with the original paper); the writing $T(x), X^i(x)$ in Equation} \eqref{eq:S-fluid-param} {clarifies their dependence as} functions of generic coordinates $x^{\alpha} =(t,x^i)$ that instead correspond to the metric $g_{\alpha \beta}$ ($\partial_{\alpha}$ stands for the derivative with respect to $x^{\alpha}$). Such reparametrization is a necessary tool for recovering a field theory that is diffeomorphism-invariant, as~opposed to field equations valid only in the Gaussian frame, simply by providing the map between the Gaussian and the arbitrary desired coordinates. Indeed, the~non-reparametrized form of \eqref{eq:S-fluid-param} would be 
\begin{linenomath}
\begin{equation}\label{eq:S-fluid-nonparam}
    S^f_{G} = \int d^4X \left[ -\frac{\sqrt{-\gamma}}{2} \left(\gamma^{00}-1 \right)\mathcal{F} + \sqrt{-\gamma}\, \,\gamma^{0i} \mathcal{F}_i \right]\,,
\end{equation}
\end{linenomath}
corresponding to a gauge fixing, where $\mathcal{F}, \mathcal{F}_i$ act as Lagrange multipliers (since their variations give the Gaussian conditions). The~reparametrized form \eqref{eq:S-fluid-param} is then uniquely obtained by requiring it to be invariant under transformations of the $x^{\alpha}$, and that, for \mbox{$x^{\alpha}\equiv(T, X^i)$}, the~expression is equivalent to \eqref{eq:S-fluid-nonparam} . 

The Hamiltonian formulation of \eqref{eq:S-fluid-param} shows how such a contribution can play the role of a physical clock for quantum gravity, when $S^f$ is adjoined to the Einstein--Hilbert action and the canonical quantization is implemented. Such formulation can be evaluated via the Arnowitt--Deser--Misner (ADM) foliation~\cite{bib:arnowitt-deser-misner-1960,bib:misner-gravitation}, {which allows one to write the line element as} 
\begin{linenomath}
\begin{equation}
        ds^2 = N^2 dt^2 - h_{ij}\, dx^i dx^j\,,
    \end{equation}
\end{linenomath}
where we label by $h_{ij}$ ($i,j$ are spatial indices) the induced metric on the identified 3d hypersurfaces $\Sigma$, and by $N$ and $N^i$ the lapse function and shift vector describing the separation in the time-like and space-like directions, respectively. The~super Hamiltonian and supermomentum contributions are
\begin{linenomath}
\begin{gather}
    \label{eq:KTfluidH}
    H^f = W^{-1} P + W W^k P_k \, ,\\
    \label{eq:KTfluidHi}
    H_i^f = P\, \partial_i T + P_k \, \partial_i X^k\,,
\end{gather}
\end{linenomath}
where $P$, $P_k$ are the momenta conjugate to $T, X^k$ and
\begin{linenomath}
\begin{gather}
    W \equiv (1- h^{jl} \partial_j T \,\partial_l T)^{-1/2} \, ,\label{eq:KTdefW}\\
    W^k \equiv h^{jl} \partial_j T \,\partial_l X^k  \,.\label{eq:KTdefWk}
\end{gather} 
\end{linenomath}

The functions \eqref{eq:KTfluidH} and \eqref{eq:KTfluidHi}, which are linear in the momenta, are added to the analogous functions $H^g, H_i^g$ of the gravitational sector; consequently, the~total constraints $H^g+H^f$ and $H^g_i+H^f_i$ must vanish because of diffeomorphism invariance~\cite{bib:cianfrani-canonicalQuantumGravity,bib:thiemann-book}. One can thus obtain a functional Schr\"odinger evolution with the time definition
\begin{linenomath}
\begin{equation}\label{eq:kuchar-torre-time-def}
    \hat{\mathcal{H}} \Psi = i\hbar\, \partial_t \Psi = i\hbar \int_{\Sigma} d^3x \frac{\delta \Psi (T,X^k,h^{jl})}{\delta T(x)} \Big|_{T=t} \,,
\end{equation}
\end{linenomath}
where $\mathcal{H} = \int_{\Sigma} d^3x \, \hat{H}^g$, i.e.,~restricting the states to the hypersurfaces where $t\equiv T$, so that $\Psi$ is still a functional of the $X^i$. The~choices $x^i \equiv X^i$ and $(t,x^i) \equiv (T,X^i)$ are also examined in the original paper~\cite{bib:kuchar-torre-1991}.

However, an~important characteristic of the Gaussian-frame method emerges at the classical level of the theory. Varying the total action with respect to the metric, it is observed that the corresponding Einsteinian equations are modified by the appearance of a source~term:
\begin{linenomath}
\begin{equation}\label{eq:Tmunu-fluid}
    T^{\alpha \beta} = \mathcal{F}\, \mathrm{U}^{\alpha} \mathrm{U}^{\beta} + \frac{1}{2} \left( \mathcal{F}^{\alpha}\,\mathrm{U}^{\beta} + \mathcal{F}^{\beta} \,\mathrm{U}^{\alpha} \right)\,,
\end{equation}
\end{linenomath}
being that $\mathrm{U}^{\alpha} = g^{\alpha \beta} \partial_{\beta} T $ and $\mathcal{F}_{\alpha} = \mathcal{F}_i \partial_{\alpha}X^i$. Thus, the Gaussian-frame terms arise as a fluid component, having four-velocity $\mathrm{U}^{\alpha}$, energy density $\mathcal{F}$, and heat flow $\mathcal{F}_{\alpha}$. The~associated energy conditions give the relation
\begin{linenomath}
\begin{equation}\label{eq:KT-energy-conditions}
    \mathcal{F} \geq 2 \sqrt{\gamma^{\alpha \beta} \mathcal{F}_{\alpha} \mathcal{F}_{\beta}} \,,
\end{equation}
\end{linenomath}
which, however, is not in general satisfied due to the arbitrariness of the Lagrange multipliers, so the fluid has a non-physical character. Actually, by implementing only the Gaussian time condition with $\mathcal{F}$ (i.e., setting $\mathcal{F}_i=0$ in \eqref{eq:S-fluid-param}), the~fluid reduces to an incoherent dust (no heat flow is present). In~this case, the~energy conditions are ensured by $\mathcal{F} \geq 0$, which can be cast as an initial condition, since $\sqrt{-g} \mathcal{F}$ is a constant of motion. We stress that this point will be differently addressed in the next~subsection.

\subsection{WKB Matter Dynamics with the Gaussian Frame~Implementation}\label{ssec:fluidWKB}

Here we briefly illustrate the procedure, discussed in~\cite{bib:maniccia-montani-2022}, by~which the kinematical variables associated with the Gaussian reference frame can provide a suitable clock for the matter sector in a quantum gravity--matter system. Indeed, unitary dynamics emerges at the next order of expansion in a Planckian parameter, where quantum gravity corrections arise. This scheme will then be applied for the computation of the modified primordial power spectrum in the next~section.

Let us consider a gravity--matter system, where the gravitational Hamiltonian is characterized by a kinetic term and a potential $V$, and~the matter component is a self-interacting scalar field $\phi$ {with potential $U_m(\phi)$}. Such choice will turn out to be suitable for the cosmological implementation discussed in Section~\ref{sec:powerspectrum}. We insert the Gaussian-frame term \eqref{eq:S-fluid-param} such that the total action reads:
\begin{linenomath}
\begin{equation}
    \label{eq:StotinADM}
    S = \int dt \int_{\Sigma} d^3x \left( \Pi^{ij} \dot{h_{ij}} +p_{\phi}\, \dot{\phi} - N ( H^g +H^m)- N^i (H_i^g+H_i^m) \right)  + S^f\,,
\end{equation}
\end{linenomath}
where
\begin{linenomath}
\begin{gather}
    \label{eq:Hgrav}
    H^g = -\frac{\hbar^2}{2M} \left( G_{ijkl}\dhij{}{ij}\dhij{}{kl} + g_{ij} \dhij{}{ij} \right) + M \, V \, , \\
    \label{eq:Higrav}
    H_i^g = \vphantom{\frac{1}{2}} 2  i\hbar\, h_{ij}\, D_k \dhij{}{kj}\, ,\\
    \label{eq:Hmatter}
    H^m = -\hbar^2 \dphisq{} +U_m \, ,\\
    \label{eq:Himatter}
    H_i^m = - (\partial_i \phi) \dphi{} \, .
\end{gather}
\end{linenomath}

In this notation, we will treat functional derivatives as ordinary partial ones{. The} term $ g_{ij} \,\partial / \partial h_{ij}$ in \eqref{eq:Hgrav} is inserted to account for a generic factor ordering (see discussion in~\cite{bib:kiefer-1991}), {and} $D_k$ in \eqref{eq:Higrav} is the 3d covariant derivative on the hypersurface $\Sigma$. Instead of the Einstein constant $\kappa = 8\pi G/c^3$, we have written in \eqref{eq:Hgrav} and \eqref{eq:Higrav} the following Planckian parameter:
\begin{linenomath}
\begin{equation}
    \label{eq:defM}
    M := \frac{c^2}{32\pi G} = \frac{c m_{Pl}^2}{4\hbar} \,,
\end{equation}
\end{linenomath}
with dimension of mass over length, which will be taken as the order parameter for the expansion. Indeed, the Planckian energy scale, representative of the gravitational sector, is typically larger with respect to the corresponding scale of the matter fields. {In principle, it would be possible to construct the WKB expansion via a dimensionless parameter, constructed as the ratio between the present one and the corresponding quantity calculated for a typical energy scale of the quantum matter. In~the case we will consider, such an energy scale corresponds to that one of the inflationary process, say, $T\simeq 10^{15}$ GeV (see Section~}\ref{sec:powerspectrum}{). However, we retain here a dimensional parameter in order to keep contact and comparison with the previous literature, e.g.,~Refs.} \cite{bib:kiefer-1991,bib:brizuela-kiefer-2016-desitter,bib:brizuela-kiefer-2016-slow-roll,bib:kiefer-2018,bib:montani-digioia-maniccia-2021,bib:maniccia-montani-2022}.

{The previous} consideration motivates a Born--Oppenheimer (B-O) separation of the wave function:
\begin{linenomath}
\begin{equation}
        \Psi \left(h_{ij}, \phi, X^{\mu} \right) = \psi \left(h_{ij}\right) \chi \left( \phi, X^{\mu} ; h_{ij}\right)
    \end{equation}
\end{linenomath}
between the gravity and matter components. After performing a WKB expansion~\cite{bib:landau-quantumMechanics} in powers of $1/M$, we have
\begin{linenomath}
\begin{equation}
    \label{eq:psiinizialeWKB}
    \Psi\left(h_{ij}, \phi, X^{\mu} \right) = e^{ \frac{i}{\hbar} \left(M S_0 +S_1 + \frac{1}{M} S_2\right) } e^{\frac{i}{\hbar} \left(Q_1 + \frac{1}{M} Q_2\right) } 
\end{equation}
\end{linenomath}
up to the order $M^{-1}$. Here, the~$S_m$ functions (at the $m$-order) account for the gravitational background, and the $Q_n$ (at the order $n$) \change{describes}{describe} the reference fluid and matter components. We stress that the first matter contribution is of the order $M^0$. Similarly to the B-O scheme, we enforce the conditions
\begin{linenomath}
    \begin{gather}
        \frac{\langle \hat{H}^m \rangle}{\langle \hat{H}^g \rangle} = \ord{M^{-1}} \,, \label{eq:condition-Hm-Hg}\\
        \dhij{Q_n}{ij} = \ord{M^{-1}}\,,\label{eq:condition-derivatives-Qn}
    \end{gather}
\end{linenomath}
where the expectation values are computed over the corresponding wave functions, due to the "fast" nature of the matter sector with respect to~gravity.

If the average backreaction of the matter degrees of freedom is negligible, both the gravitational and total constraints are satisfied:
\begin{linenomath}
\begin{gather}
    \label{eq:vincologravH}
    \hat{H}^g \,\psi (h_{ij}) =0 \, , \\
    \label{eq:vincologravHi}
    \hat{H}_i^g \,\psi(h_{ij}) =0 \, ,\\
    \label{eq:vincolototH}
    (\hat{H}^g + \hat{H}^m + \hat{H}^f ) \Psi (h_{ij},\phi, X^{\mu}) =0 \,,\\
    \label{eq:vincolototHi} 
    (\hat{H}_i^g + \hat{H}_i^m + \hat{H}_i^f ) \Psi (h_{ij},\phi, X^{\mu}) =0\,,
\end{gather}
\end{linenomath}
where we consider also the supermomentum constraints for generality.
By substituting the ansatz \eqref{eq:psiinizialeWKB} in the constraints, with~the explicit forms \eqref{eq:KTfluidH}, \eqref{eq:KTfluidHi}, and \eqref{eq:Hgrav}--\eqref{eq:Himatter}, the~dynamics can be analyzed order by order (we refer to the original paper~\cite{bib:maniccia-montani-2022} for the explicit computation). 

At the Planckian order $M$, one obtains the classical Hamilton--Jacobi (H-J) equation for the gravitational function $S_0$:
\begin{linenomath}
\begin{equation}\label{eq:hamilton-jacobi}
        \frac{1}{2} G_{ijkl} \frac{\partial S_0}{\partial h_{ij}} \frac{\partial S_0}{\partial h_{kl}} + V =0 \,,
    \end{equation}
\end{linenomath}
together with its diffeomorphism invariance condition. A~crucial point must instead be discussed at the order $M^0$: the gravitational constraints \eqref{eq:vincologravH} and \eqref{eq:vincologravHi} allow one to solve for $S_1$, and after~substituting the solutions $S_0$ and $S_1$ into \eqref{eq:vincolototH} and \eqref{eq:vincolototHi}, the~remaining equations for the matter sector are
\begin{linenomath}
\begin{gather}
    \label{eq:HtotOrdM0}
    \left(-2i\hbar \dphisq{Q_1} +U_m -W^{-1} \frac{\partial Q_1}{\partial T}-W W^k \frac{\partial Q_1}{\partial X^k}\right) e^{\frac{i}{\hbar}Q_1} =0 \,,\\
    \label{eq:HitotOrdM0}
    \left(-2h_{ij}\, D_k \dhij{S_1}{kj} -i\hbar^{-1} (\partial_i \phi) \dphi{Q_1} - (\partial_i T) \frac{\partial Q_1}{\partial T} - (\partial_i X^k) \frac{\partial Q_1}{\partial X^k} \right) e^{\frac{i}{\hbar}Q_1} =0 \, .
\end{gather}
\end{linenomath}

Such expressions require further attention for their physical interpretation. Following from the linearity of $H^f$ and $H^f_i$ in the momenta $P$ and $P_k$, a~suitable time parameter can be naturally introduced as
\begin{linenomath}
\begin{equation}
\label{eq:fluid-def-time}
    i \hbar \dtau{}= i\hbar \int d^3x \left[ N\left( W^{-1} \frac{\partial}{\partial T} +W W^k \frac{\partial}{\partial X^k} \right) +N^i\left( (\partial_i T)\frac{\partial}{\partial T} +(\partial_i X^k) \frac{\partial}{\partial X^k} \right) \right]\,.
\end{equation}
\end{linenomath}

Then, the~linear combination of Equations~\eqref{eq:HtotOrdM0} and  \eqref{eq:HitotOrdM0} with coefficients $N$ and $N^i$, respectively, takes the form
\begin{linenomath}
\begin{equation}
\label{eq:fluid-schrod-eq-M0}
    i \hbar \dtau{\chi_0} = \hat{\mathcal{H}}^m \chi_0 = \int d^3x \left( N\hat{H}^m + N^i \hat{H}^m_i\right) \chi_0\,,
\end{equation}
\end{linenomath}
where we label $\chi_0 = e^{\frac{i}{\hbar} Q_1}$. In~other words, reading the reference fluid clock induced by the definition \eqref{eq:fluid-def-time}, we observe at this WKB order functional Schr\"odinger dynamics of the quantum matter field $\phi$ on the gravitational background. In~this limit, the~resulting dynamics \change{correspond}{corresponds} to QFT on curved~spacetime.

The quantum gravity's influence on the matter sector emerges at the next order, $M^{-1}$. Proceeding in a similar way, one obtains the following equation for $\chi_1 = e^{\frac{i}{\hbar} \left(Q_1 +\frac{1}{M} Q_2\right)}$:
\begin{linenomath}
\begin{equation}\label{eq:final-fluid}
   i\hbar \dtau{\chi_1} = \hat{\mathcal{H}}^m \chi_1 + \int d^3 x \left[ N G_{ijkl} \dhij{S_0}{ij}\left( -i\hbar\dhij{}{kl} \right) - 2N^i h_{ij}\, D_k \left( -i\hbar \dhij{}{kj}\right) \right] \chi_1\,,
\end{equation}
\end{linenomath}
where the additional contributions with respect to the matter Hamiltonian are quantum gravity corrections. These modifications are unitary due to the real nature of the function $S_0$ and the presence of the conjugate momenta with respect to the induced metric, and their smallness is assured by the hypothesis \eqref{eq:condition-derivatives-Qn} (see  \cite{bib:maniccia-montani-2022}). Thus, the clock defined by \eqref{eq:fluid-def-time} is a physical clock for the matter sector in the WKB scheme truncated at the order $M^{-1}$.

\section{Calculation of the Inflationary~Spectrum}\label{sec:powerspectrum}

Following the model introduced in the previous section, we now turn to the question of how the power spectrum associated with inflationary perturbations is affected by the quantum gravity~corrections.

\subsection{Perturbations of the~Model}\label{ssec:disc-perturbations}
{Before facing the analysis of the generation of primordial perturbations during the inflationary dynamics of the Universe, and when~studying how the quantum gravity corrections can affect the associated power spectrum, it is worth stressing some key differences between the present analysis and other similar approaches, as~in} \cite{bib:gundlach-1993,bib:brizuela-kiefer-2016-slow-roll}. 

{In our formulation, apart from the WKB expansion in the Planckian parameter $M$, we are addressing a B-O separation between the "slow" gravitational component and the "fast" matter contribution, with the~latter including also the fluid's presence.
This separation is justified by virtue of a corresponding scale separation between the energy of the quantum matter dynamics, say in the order of the matter Hamiltonian spectrum, and~that one of the Planck order, at~which the gravity quantization is expected to manifest itself.
In view of the adopted B-O approximation we are implementing, the~backreaction of the quantum matter on the gravitational background is implicitly negligible. In~other words, quantum corrections of the gravitational dynamics are clearly present (as implied by the function $S_1$ in Equation} \eqref{eq:psiinizialeWKB}{\change{and}{,} associated with a quantum amplitude for the background metric), but~their existence has to be regarded as independent of the matter's dynamics.}

{This point of view has been clearly elucidated in}  \cite{bib:maniccia-montani-antonini-2023}, {where a critical re-analysis of the original formulation} \cite{bib:vilenkin-1989}, {and hence,} Ref.~\cite{bib:kiefer-1991}, {has been developed. 
There, limiting the attention up to the zero order in the parameter $M$, the~quantum gravity component has been expressed in terms of gravitons on the vacuum Bianchi I background. This way, the~WKB formulation of the gravitational field takes the form of a purely classical background on which a slow quantum graviton field lives, as~referred to by independent degrees of freedom. 
This graviton contribution is independent, due to the B-O separation, from~the quantum matter dynamics, thereby reinforcing the previous statement. 
If implemented to the isotropic Universe we will consider below, this formulation would also imply the presence of scalar perturbations of the metric, represented by independent degrees of freedom, and clearly, not affected by the scalar field fluctuations.}

{In the following analysis, although~developed in the presence of quantum gravity corrections, we will refer to the scalar field only; such case is equivalent to the study of a free massless scalar field fluctuating on a de Sitter background. The~classical energy contribution of the scalar field will be identified with the cosmological constant term (i.e., the gap between the false and true vacuum energy density} \cite{bib:kolb-turner,bib:montani-primordialcosmology}). {The inhomogeneous fluctuation of this field will be treated as an independent degree of freedom living on the expanding de Sitter space and whose fluctuations are responsible for the emergence of a scalar perturbation spectrum.}

\subsection{The Inflaton~Field}\label{ssec:inflaton}
The theory of inflation {postulates an early period of exponentially accelerated expansion of the Universe, motivating its primordial inhomogeneities as emerging from the vacuum fluctuations of a scalar field, the~so-called inflaton field. One of the most remarkable results of such mechanism is the ability to explain the flatness problem of the Universe} \cite{bib:weinberg,bib:montani-primordialcosmology,bib:brandenberger-book-2004,bib:peter-uzan-PC}.
{A schematic formulation of this framework is studied by considering as a background a spatially flat universe (any curvature is damped by the exponential expansion), with~a scalar field living on top. More specifically, one should consider a Friedmann--Lemaitre--Robertson--Walker (FLRW) model with line element}:
\begin{linenomath}
\begin{equation}\label{eq:FLRW_line-element}
      ds^2 = -N^2(t) \;dt^2 + a^2(t)\left(dx^2 + dy^2 + dz^2\right)\,,
    \end{equation}
\end{linenomath}
{ $a$ being the cosmic scale factor (here we use the opposite signature with respect to} \eqref{eq:S-fluid-param} {in Section}~\ref{sec:fluid} {for easier comparison with existing literature), and~it inserts the inflaton contribution as a minimally coupled scalar field $\phi$ with potential $U(\phi)$. Then, small perturbations are introduced, in~general, both for the metric and for the inflaton, which give rise to scalar and tensor fluctuations (the detailed Hamiltonian formulation of such approach can be found, for~example, in} \cite{bib:brizuela-kiefer-2016-desitter,bib:brizuela-kiefer-2016-slow-roll}). {Following the discussion presented in Section~}\ref{ssec:disc-perturbations}, {we will now focus on the fluctuations of the scalar field only over the FLRW background, i.e.,~by variation of the action with respect to those variables.}

The fluctuations $\delta \phi$ of the inflaton field can be described in a gauge-invariant way via the Mukhanov--Sasaki (M-S) variable $v$ 
\cite{bib:mukhanov-1985,bib:sasaki-1986,bib:mukhanov-1988} (see also the discussion in~\cite{bib:gundlach-1993}) {defined as} 
\begin{linenomath}
\begin{equation}\label{def:v-MS}
        v := a \varphi = a \,\delta \phi\,.\note{adjusted name of the variable}
    \end{equation}
\end{linenomath}

{We here stress that addressing the B-O separation discussed above does not alter the gauge invariance of the perturbation theory. In~fact, in~the limit in which the backreaction on the metric scalar perturbation is neglected, the~M-S variable} \cite{bib:mukhanov-1985} {simply reduces to the inhomogeneous scalar field $\phi$ of our study times the cosmic scale factor, as~in} \eqref{def:v-MS}, {and its gauge invariance is immediately recovered.}

{The evolution of the inflaton fluctuations is responsible for the formation of primordial structures in the Universe. To~analyze their behavior, let us consider modes with physical wavelength $\lambda_{phys} \equiv a(t) \lambda_0$,  $\lambda_0$ being the comoving wavelength. It is useful to compare this quantity with the so-called Hubble radius} (or micro-physics horizon) $\Hubble^{-1} = a/\dot{a}$, that for any given time is the inverse of the Hubble parameter (using $c=1$). This horizon represents the scale separating the gravity-dominated regime from the quantum one: the first happens for modes with physical wavelength  such that $\lambda_{phys} \gg \Hubble^{-1}$, and the second is the case for $\lambda_{phys} \ll \Hubble^{-1}$. {It can be shown that,} during the period of accelerated expansion predicted by the theory, the~Hubble radius is constant in the physical coordinates, and $\lambda_{phys}$ exponentially increases \cite{bib:montani-primordialcosmology,bib:peter-uzan-PC}. Thus, the~quantum fluctuations emerge at early times within the micro-physical scales (i.e., for $\lambda_{phys} \ll \Hubble^{-1}$), rapidly expand going outside the horizon, and~propagate until they re-enter the Hubble radius at later times (when inflation is over, the~behavior is opposite, since $\Hubble^{-1}$ grows faster than the $\lambda_{phys}$) \cite{bib:weinberg,bib:brandenberger-book-2004}.

Using {the gauge-invariant formalism via the M-S variable} \eqref{def:v-MS}, {it is possible to} compute the power spectrum $\mathcal{P}_v(k)$, where $k$ specifies the wavenumber of each Fourier mode {associated with the inflaton perturbations (see also} \cite{bib:martin-olmedo-2016-LQC,bib:olmedo-singh-2020,bib:gielen-2022-spectrum-bounce,bib:kiefer-tatevik-2022,bib:cheng-2022,bib:bortolotti-2022} {for investigations of such a spectrum in different cosmological settings). However, to~investigate the evolution of the primordial Universe, it is more convenient to work with the spectrum associated with the} comoving curvature perturbation $\zeta$ {(which is the one leaving its fingerprint on the cosmic microwave background radiation)} \cite{bib:martin-vennin-peter-2012}: indeed, $\zeta$ is constant (i.e., it freezes) for all the time in which the perturbations are outside the horizon; therefore, one only needs to compute its spectrum  at the end of inflation~\cite{bib:montani-primordialcosmology,bib:peter-uzan-PC} . In~the primordial era of our interest, the~two quantities $\zeta$ and $v$ are directly related by
\begin{linenomath}
\begin{equation}\label{def:curvature-perturbation}
        \zeta = \sqrt{\frac{4\pi G}{\epsilon}}\frac{v}{a},
    \end{equation}
\end{linenomath}
with $\epsilon = -\dot{\textrm{H}}/\textrm{H}^2$ being the first slow-roll parameter. Therefore, in~the following, we will focus on the dynamics of the M-S variable $v$ and only at the end use \eqref{def:curvature-perturbation} to compute the invariant power~spectrum. 

Upon decomposition in Fourier modes $v_{\mathbf{k}}$ and assuming Gaussian probability distributions for the quantum amplitudes associated with each $v_{\mathbf{k}}$ \cite{bib:martin-vennin-peter-2012}, all the {relevant} properties of the inflationary perturbations are contained in the two-point correlation function:
\begin{linenomath}
\begin{equation}\label{def:correl-function}
    \Xi(\mathbf{r}) := \langle 0 | \hat{v}(\eta,\mathbf{x}) \hat{v}(\eta, \mathbf{x}+\mathbf{r})|0\rangle \,,
\end{equation}
\end{linenomath}
where $|0\rangle$ is the vacuum state of the inflaton field. In~\eqref{def:correl-function}, the~expectation value implies integration over $\mathbf{k}$-modes, which can be carried out given the expression~\cite{bib:martin-vennin-peter-2012}
\begin{linenomath}
\begin{equation}\label{eq:correl-function-amplitude}
    \Xi(\mathbf{r})= \frac{1}{(2\pi)^3} \int d\mathbf{p} \;e^{-i \mathbf{p}\cdot \mathbf{r}} |f_{\mathbf{p}}|^2 = \frac{1}{2\pi^2} \int_0^{+\infty} \frac{dp}{p} \frac{sin(pr)}{pr} p^3 |f_p|^2 \,.
\end{equation}
\end{linenomath}

Here, $f_{\mathbf{p}}$ is the mode function associated with the scalar perturbations, and~from \eqref{eq:correl-function-amplitude}, the power spectrum is defined as
\begin{linenomath}
\begin{equation}\label{def:power-spectrum}
    \mathcal{P}_{v} (k) = \frac{k^3}{2\pi^2} |f_k|^2\,,
\end{equation}
\end{linenomath}
i.e., the Fourier amplitude of $\Xi(0)$ per unit logarithmic interval. {As mentioned} above, this quantity is {then} evaluated in the super-Hubble limit $k/(a\Hubble) \ll 1$, when the perturbations essentially freeze.
We stress that the vacuum state in \eqref{def:correl-function} must be selected as the one corresponding to the ground level of the scalar field Hamiltonian in the limit $k/(a\Hubble) \rightarrow \infty$ (or equivalently $\lambda_{phys} \ll \Hubble^{-1}$), also known as the Bunch--Davies vacuum~\cite{bib:martin-vennin-peter-2012,bib:brizuela-kiefer-2016-desitter,bib:weinberg}. We will impose this requirement on the modified wave functional dictated by the model in Section~\ref{ssec:applic-alpha}.

In the following, we will compute \eqref{def:power-spectrum} in the specific case where the inflaton field follows modified quantum dynamics, as~described by Equation~\eqref{eq:final-fluid}.

\subsection{Perturbation Spectrum in the de Sitter~Phase}\label{ssec:applic-alpha}
 
During the accelerated expansion of inflation, of~particular interest is the slow-rolling phase, where the inflaton can be approximately described as a free massless scalar field (the almost constant potential acts as a cosmological term) \cite{bib:montani-primordialcosmology}. In~the following, we will consider an exact de Sitter phase; thus, the slow-rolling parameter $\epsilon$ is neglected. The~analysis of quantum gravity's effects on the inflationary spectrum is achieved by considering the fluctuations of the scalar field over a quasi-classical background, {expressed by a FLRW model with line element} \eqref{eq:FLRW_line-element}.

Instead of the general action \eqref{eq:StotinADM}, the~considered case can be studied in the minisuperspace formalism (the supermomentum contributions are identically vanishing due to the homogeneity of the {background} model). The~(non-trivial) relevant constraint is thus the superHamiltonian, which takes the form
\begin{linenomath}
\begin{equation}\label{eq:superHtot-applic}
    H_{tot} = \frac{\hbar^2}{48Ma^2}\partial_a\left(a\partial_a\right) + 4 M\Lambda a^3- i\hbar\partial_T +\frac{1}{2a}\sum_{\mathbf{k}}\left( -\hbar^2 \partial^2_{v_{\mathbf{k}}} + \omega_{k}^2 v_\mathbf{k}^2 \right)\,.
    \end{equation}
\end{linenomath}
where we implemented the Laplace--Beltrami factor ordering. Here, the positive cosmological constant $\Lambda$ replaces $U_m$ in \eqref{eq:Hmatter}. The~term $-i\hbar \partial_T$, that is, the momentum associated with the Gaussian time $T$, is the only surviving contribution from the insertion of $S^f$ \eqref{eq:S-fluid-param} due to homogeneity. The~last two terms in \eqref{eq:superHtot-applic} are associated with the inflaton field fluctuations, where the $v_{\mathbf{k}}$ correspond to {the modes in the Fourier space of the gauge-invariant M-S variable}~\eqref{def:v-MS}; in the considered case of scalar perturbations over a FLRW background, the~$v_{\mathbf{k}}$-modes behave as time-dependent harmonic oscillators~\cite{bib:langlois-1994,bib:brizuela-kiefer-2016-desitter,bib:brizuela-kiefer-2016-slow-roll,bib:giesel-2020,bib:venturi-2021}, where the frequency depends on the wavenumber modulus only:
\begin{linenomath}    \begin{equation}\label{eq:def-omega-TDHA}
    \omega_{k}^2 = k^2 - \frac{a^2}{N^2}\left(\dot{\Hubble} - \Hubble \frac{\dot{N}}{N} + 2\Hubble^2 \right)\,.
\end{equation}
\end{linenomath}

The WDW constraint corresponds to the vanishing of the operator \eqref{eq:superHtot-applic} applied to the total system wave function $\Psi (a, T, v_{\mathbf{k}})$.
For convenience, we implement the logarithmic scale factor,
\begin{linenomath}
\begin{equation}\label{def-alpha}
    \alpha:= \ln \left(\frac{a}{a_0}\right)\,,
\end{equation}
\end{linenomath}
such that the global WDW equation reads
\begin{linenomath}
\begin{equation}\label{eq:superHtot-applic-alpha}
    i\hbar\partial_T \Psi = a_0^{-1}e^{-\alpha} \left[\frac{\hbar^2 }{48M} \frac{1}{a_0^2 e^{2\alpha}}\partial_\alpha^2 + 4a_0^4e^{4\alpha}\Lambda M +\frac{1}{2} \sum_{\mathbf{k}}\left( -\hbar^2 \partial^2_{v_{\mathbf{k}}} + \omega_{k}^2 v_\mathbf{k}^2 \right)\right]\Psi\,.
\end{equation} 
\end{linenomath}

Let us now consider a single Fourier mode identified by a wave number $\mathbf{k}$. Following the scheme discussed above, for~each independent mode, the ansatz is taken as
\begin{linenomath}
\begin{equation}
         \Psi_\mathbf{k}(\alpha, T, v_{\mathbf{k}}) = \psi_\mathbf{k}(\alpha)\;\chi_\mathbf{k}(\alpha, T, v_\mathbf{k})\,,
    \end{equation}
\end{linenomath}
and then WKB expanded as in \eqref{eq:psiinizialeWKB}, obtaining
\begin{linenomath}
    \begin{gather}
           \psi_{\mathbf{k}}(\alpha) =e^{\frac{i}{\hbar}\left[MS_0(\alpha) + S_1(\alpha) + M^{-1} S_2(\alpha)\right]}\,,\\
           \chi_{\mathbf{k}}(\alpha, T, v_\mathbf{k}) = e^{\frac{i}{\hbar}\left[Q_1(\alpha, T, v_\mathbf{k}) + M^{-1} Q_2(\alpha, T, v_\mathbf{k}))\right]}\,.
    \end{gather}
\end{linenomath}

Upon substitution into \eqref{eq:superHtot-applic-alpha}, the~solutions for the gravitational sector are readily obtained at the three orders:
\begin{linenomath}
    \begin{gather}
        S_0  (\alpha) = -8\sqrt{\frac{\Lambda}{3}}a_0^3\left(e^{3\alpha} - e^{3\alpha_0}\right)\,,\label{eq:sol-S0}\\
        S_1(\alpha) = i\hbar\frac{3}{2}(\alpha - \alpha_0)\,,\label{eq:sol-S1}\\
        S_2(\alpha) = \frac{\hbar^2}{64}\sqrt{\frac{3}{\Lambda}}a_0^{-3} \left( e^{-3\alpha} - e^{-3\alpha_0} \right)\,.
    \end{gather}
\end{linenomath}

Here, $S_0$ solves the H-J equation and so corresponds to the classical limit of the gravitational component, and the next order functions, $S_1$ and $ S_2$, account for quantum gravity~effects.

The equation for the quantum matter wave function at the first order $M^0$ can be expressed in a clearer form in conformal time $\eta$ (choosing {$N= a_0 \,e^{\alpha}$}), which is related to the Gaussian time constraint via $T'(\eta) = a_0\exp\left({\alpha(\eta)}\right)$, obtaining
\begin{linenomath}
\begin{equation}\label{eq:TDharmonic-oscillator}
       i\hbar \partial_\eta \chi_{\mathbf{k}}^{(0)} = \left(- \frac{\hbar^2}{2}\partial_{v_{\mathbf{k}}}^2 + \frac{1}{2}\omega_k^2(\eta)  v_{\mathbf{k}}^2 \right)\chi_{\mathbf{k}}^{(0)}\,. 
   \end{equation} 
\end{linenomath}

The time-dependent harmonic oscillator system can be exactly solved by implementing the so-called Lewis--Riesenfeld method introduced in~\cite{bib:lewis-1967,bib:lewis-1968,bib:lewis-riesenfeld-1969,bib:pedrosa-1997}, which is described in appendix \ref{appendix:Lewis}.  The~wave function admits a general representation of the form \eqref{eq:basis-chi0-invariant-eigenstates}, where the functions $\delta_{n,k}$ and $\rho_k$ are defined in \eqref{eq:delta-nk-invariant} and \eqref{eq:rho-solution}, respectively. The~arbitrary coefficients in those expressions are set by imposing suitable initial conditions. In~this specific cosmological setting, we make use of the Bunch--Davies vacuum state requirement~\cite{bib:weinberg,bib:martin-vennin-peter-2012}: the state must correspond to the Minkowskian vacuum in the limit $\eta\to-\infty$ (that is, when the inflaton wavelength is small compared to the curvature of the universe). This condition is satisfied if
\begin{linenomath}
    \begin{gather}
        \rho_k(\eta) \xrightarrow{\eta\to-\infty} k^{-1/2}\,,\\
        c_{n,k} = \delta_{0,k}\,\label{eq:requirement-cnk}
    \end{gather}
\end{linenomath}
where \eqref{eq:requirement-cnk} stems from the observation that the $n=0$ eigenvalue of the invariant \eqref{eq:invariant-I} corresponds, for~a fixed time, to~the lowest-energy state of the oscillator. For~the specific $\rho_k$ function \eqref{eq:rho-solution}, its coefficients must be $A=B=\gamma_1=1$, so that
\begin{linenomath}
\begin{equation}
        \rho_k(\eta) = \sqrt{\frac{1}{k} + \frac{1}{\eta^2 k^3}}
    \end{equation}
\end{linenomath}
satisfies the required limit. Then, by substituting it into \eqref{eq:delta-nk-invariant}, the~$\delta_{n,k}$ functions are found to be
\begin{linenomath}
\begin{equation}
        \delta_{n,k} = -\left(n+\frac{1}{2}\right)\int d\eta \frac{1}{\rho_k^2(\eta)} = -\left(n+\frac{1}{2}\right)\left(\eta k -\arctan(\eta  k) + c\right)\,.
    \end{equation}
\end{linenomath}

Finally, the~solution to Equation~\eqref{eq:TDharmonic-oscillator} satisfying the Bunch--Davies condition is:
\begin{adjustwidth}{-\extralength}{0cm}
\begin{equation}\label{eq:applic-sol-chi0-BD}
     ^{BD} {\chi}_{\mathbf{k}}^{(0)}(\eta, v_{\mathbf{k}}) = \exp\left[{-\frac{i}{2}\left(\eta k - \arctan(\eta k)\right)}\right] 
        \left(\frac{k^3}{\pi\hbar \left(\frac{1}{\eta ^2}+k^2\right)}\right)^{\frac{1}{4}} \exp \left[\frac{i}{2\hbar}\left(-\frac{1}{\eta ^3 \left(\frac{1}{\eta ^2}+k^2\right)} + i \frac{k^3}{\frac{1}{\eta ^2}+k^2}\right) v_{\mathbf{k}}^2\right]
\end{equation}
\end{adjustwidth}

We can now focus on the next order $M^{-1}$, where, due to the quantum gravity corrections, the~dynamics is no longer that of a time-dependent oscillator:
\begin{linenomath}    \begin{equation}\label{eq:applic-chi1-conS0}
        i\hbar \partial_\eta \chi_{\mathbf{k}}^{(1)} =\left[\frac{i \hbar}{24}\frac{1}{a_0^2 e^{2\alpha}}(\partial_\alpha S_0)\partial_\alpha - \frac{\hbar^2}{2}\partial_{v_{\mathbf{k}}}^2 + \frac{1}{2}\omega_k^2 v_{\mathbf{k}}^2 \right]\chi_{\mathbf{k}}^{(1)}\,.
    \end{equation}
\end{linenomath}

By substituting \eqref{eq:sol-S0} and the classical background solution $a_0e^\alpha (\eta) = -\sqrt{\frac{3}{\Lambda}}\frac{1}{\eta}$, Equation~\eqref{eq:applic-chi1-conS0} becomes
\begin{linenomath}
\begin{equation}\label{eq:applic-chi1-coneta}
        i\hbar \partial_\eta \chi_{\mathbf{k}}^{(1)}(\alpha, \eta, v_{\mathbf{k}})  = \left[ \frac{i \hbar}{\eta} \partial_\alpha - \frac{\hbar^2}{2}\partial_{v_{\mathbf{k}}}^2 + \frac{1}{2}\omega_k^2(\eta) v_{\mathbf{k}}^2 \right]\chi_{\mathbf{k}}^{(1)}(\alpha, \eta, v_{\mathbf{k}})\,.
    \end{equation}
\end{linenomath}

We investigate the class of separable solutions of the form
\begin{linenomath}
\begin{equation}\label{eq:applic-chi1-theta-gamma}
        \chi^{(1)}_{\mathbf{k}}(\alpha, \eta, v_{\mathbf{k}}) = \theta(\alpha)\,\Gamma_{\mathbf{k}}(\eta, v_{\mathbf{k}})\,,
    \end{equation}
\end{linenomath}
where we remark that the (quantum) degree of freedom $\alpha$ is in principle independent from the chosen conformal time $\eta$, and~the classical relation only stands in the appropriate low-energy limit.
Then, Equation~\eqref{eq:applic-chi1-coneta} is solved for
\begin{linenomath}
\begin{gather}
    -i\hbar \partial_{\alpha}\theta(\alpha) = \lambda\theta(\alpha)\,, \label{eq:eq-applic-theta}\\
    i\hbar\partial_\eta \Gamma_\mathbf{k}(\eta,v_\mathbf{k}) = \left(-\frac{\hbar^{2}}{2} \partial_{v_\mathbf{k}}^2 + \frac{1}{2}\omega_{k}^2(\eta) v_\mathbf{k}^2  -\frac{\lambda}{\eta} \right)\Gamma_\mathbf{k}(\eta,v_\mathbf{k})\,,\label{eq:applic-gamma}
\end{gather}
\end{linenomath}
where the constant $\lambda$ identifies the family of solutions of \eqref{eq:eq-applic-theta}, which gives the eigenvalues of the momentum associated with $\alpha$ and so to the scale factor $a$.
Equation~\eqref{eq:applic-gamma} can be solved via another suitable rescaling, $\Gamma_{\mathbf{k}}(\eta,v_\mathbf{k}) = \exp\left[\frac{i}{\hbar}\lambda \log(-\eta)\right] \Tilde{\Gamma}_\mathbf{k}(\eta,v_{\mathbf{k}})$, which absorbs the $\lambda$-factor and maps it into an equation of the form \eqref{eq:TDharmonic-oscillator} for $\Tilde{\Gamma}_{\mathbf{k}}$, i.e.,~the usual time-dependent harmonic oscillator. Therefore, the~function $\Tilde{\Gamma}_\mathbf{k}$ coincides with the $\chi_{\mathbf{k}}^{(0)}$ of the previous order, and the $\Gamma_{\mathbf{k}}$ is readily obtained from the rescaling above. \change{by}{By} putting together the solutions of~\eqref{eq:eq-applic-theta} and  \eqref{eq:applic-gamma}, we can write the complete matter wave function \eqref{eq:applic-chi1-theta-gamma} as
\begin{linenomath}
\begin{equation}\label{eq:applic-chi1-solution-noaverage}
      \chi^{(1)}_{\mathbf{k}}(\alpha, \eta, v_{\mathbf{k}}) = \theta_{p_{\alpha}} (\alpha)\; e^{\frac{i}{\hbar}p_{\alpha}\, log(-\eta)}\, \chi_{\mathbf{k}}^{(0)}(\eta,v_{\mathbf{k}}) \,,
    \end{equation}
\end{linenomath}
which can then be implemented to analyze the quantum-gravity corrected power~spectrum.

However, before~that computation, we stress one important remark of this approach. The~requirement \eqref{eq:condition-derivatives-Qn} imposed in Section~\ref{ssec:fluidWKB} due to the B-O approximation scheme translates, in~this specific minisuperspace setting, to~$|p_{\alpha}|< 1/M$. Therefore, one must consider for \eqref{eq:applic-chi1-solution-noaverage} a convolution over the suitable values of the momentum $p_{\alpha}$
\begin{linenomath}    \begin{equation}\label{eq:chi1-weight-p-alfa}
        {\chi}^{(1)}_{\mathbf{k}}(\alpha, \eta, v_{\mathbf{k}}) = \chi^{(0)}_{\mathbf{k}}(\eta, v_{\mathbf{k}}) \int d p_\alpha g(p_\alpha) \theta_{p_\alpha}(\alpha) e^{\frac{i}{\hbar}\log(-\eta) p_\alpha}\,,
    \end{equation}
\end{linenomath}
with $g(p_{\alpha})$ being a generic distribution. More specifically, choosing a Gaussian weight with deviation $\sigma$ and zero mean value
\begin{linenomath}    \begin{equation}\label{gaussian-weight}
        g(p_{\alpha}) = \frac{1}{(\sqrt{2\pi}\sigma)^{1/2}} e^{- \frac{p_{\alpha}^2}{4\sigma^2}}\,,
    \end{equation}
\end{linenomath}
the matter wave function modified by quantum gravity corrections ends up as
\begin{linenomath}
\begin{equation} \label{eq:chi-gaussian-weight-integrated}           
    \chi^{(1)}_{\mathbf{k},Gauss}(\alpha,\eta,v_{\mathbf{k}}) = \chi^{(0)}_{\mathbf{k}}(\eta,v_{\mathbf{k}}) \left\{ (8\pi\sigma^2)^{1/4}  \exp\left[{-\frac{\sigma ^2}{\hbar^2} \left(\alpha +\log(-\eta)\right)^2 }\right] \right\}.
    \end{equation}
\end{linenomath}

We observe that the effect of the quantum gravity corrections  has clearly factorized, an~aspect which will deeply impact the result of the power spectrum analysis. {Indeed, the~obtained wave function shall be considered as the "new" vacuum state in order to derive the primordial power spectrum for the order $M^{-1}$ of the prescribed theory, i.e.,~modified by quantum gravity effects. However, since the modification affecting the wave function}~\eqref{eq:chi-gaussian-weight-integrated} {takes the form of a time factor only, such a spectrum will coincide with the previous order result, which is computed with the wave function} \eqref{eq:applic-sol-chi0-BD} {in the absence of quantum gravitational corrections.} 

{At this stage,} the wave function $\chi^{(1)}$ retains remarkable dependence on the quantum variable $\alpha$ in the proposed paradigm, a~property which has to be carefully addressed when studying phenomenological implications. Following the considerations in~\cite{bib:maniccia-montani-2022}, we consider an "averaged" wave function in the form of 
\begin{linenomath}
\begin{equation}\label{eq:chi-mediated-A}
            \Bar{\chi}(\eta, v_{\mathbf{k}}) = \int d\alpha |A|^2(\alpha) \,\chi (\alpha,\eta,v_{\mathbf{k}}) 
    \end{equation}
\end{linenomath}
where $A = e^{i S_1/\hbar}$ is the (quantum) amplitude coming from the lowest-order quantum gravitational component. This choice corresponds to averaging on the quasi-classical gravitational probability density, which in the selected minisuperspace is associated with the logarithmic scale factor $\alpha$ only. {It is worth stressing that weighting the matter wave function on the WKB amplitude of the gravitational field is, on~the present level, a~purely phenomenological procedure. In~fact, it is clear that such a wave function can in principle no longer satisfy the Schr\"odinger equation} \eqref{eq:applic-chi1-conS0}. {Nonetheless, the applicability of the analysis in} \cite{bib:maniccia-montani-antonini-2023} is reliable, {where it has been shown that such a calibrated wave function is actually a solution of the Schr\"odinger equation when suitable gauge invariance of the B-O procedure is taken into account.}

Upon substitution of \eqref{eq:chi-gaussian-weight-integrated} and \eqref{eq:sol-S1} into Equation~\eqref{eq:chi-mediated-A}, the~averaged wave function for each mode becomes
\begin{linenomath}
\begin{equation}
        \Bar{\chi}^{(1)}_{\mathbf{k},Gauss} (\eta, v_{\mathbf{k}}) = \chi_{\mathbf{k}}^{(0)}(\eta, v_{\mathbf{k}}) \left[\hbar\left(\frac{8 \pi^3}{\sigma^2}\right)^{\frac{1}{4}} (-\eta)^{3} \exp\left(\frac{9 \hbar^2}{4 \sigma ^2}\right)\right].
    \end{equation}
\end{linenomath}

Requiring normalization over the possible $v_{\mathbf{k}}$ values, {i.e., dividing by the wave function integrated on such variables,} the term in squared brackets (which \add{depends only on time and on the specific form of the weight} \eqref{gaussian-weight}) {clearly} factors out of the integration{. Therefore, we have for the averaged and normalized wave function}
\begin{linenomath}
\begin{equation}\label{eq:chi-mediated-coincides-chi0}
        \Bar{\chi}^{(1)}_{\mathbf{k},Gauss} \xrightarrow{\textit{integration over } \alpha} \chi_{\mathbf{k}}^{(0)} (\eta, v_{\mathbf{k}}),
    \end{equation}
\end{linenomath}
{namely, we recover the previous order state.}  

{Therefore, we now proceed to the computation of the inflationary power spectrum in the described setting, by~computing} the two-point correlation function of the M-S variable on the the Bunch--Davies state \eqref{eq:applic-sol-chi0-BD}. For~convenience, we rewrite $^{BD} \chi_{\mathbf{k}}^{(0)}$ in the following way:
\begin{linenomath}
\begin{equation}\label{eq:BD-state-Omega}
        ^{BD} \chi_{\mathbf{k}}^{(0)} (\eta, v_{\mathbf{k}}) = N_k(\eta) \exp\left(i\delta_{0,k}(\eta)-\Omega_k(\eta)v_{\mathbf{k}}^2\right)\,,
    \end{equation}
\end{linenomath}
where
\begin{linenomath}
    \begin{gather}
        \Omega_k(\eta) := \frac{1}{2\hbar}\left(\frac{i}{\eta ^3 \left(\frac{1}{\eta ^2}+k^2\right)} +  \frac{k^3}{\frac{1}{\eta ^2}+k^2}\right)\,,\label{eq:def-Omega-BDstate}\\
        N_k(\eta) := \left(\frac{2}{\pi}\Re(\Omega_k) \right)^{1/4} = \left(\frac{k^3}{\pi\hbar \left(\frac{1}{\eta ^2}+k^2\right)}\right)^{\frac{1}{4}},
    \end{gather}
\end{linenomath}
and $\Re(\cdot)$ isolates the real part. In~the following, we also isolate the real and imaginary parts of the (complex) variable $v_{\mathbf{k}}$ as
\begin{linenomath}
\begin{equation}
        v_{\mathbf{k}} = \frac{1}{\sqrt{2}}(v_{\mathbf{k}}^R + i v_{\mathbf{k}}^I)
    \end{equation}
\end{linenomath}
for the computation of the correlation function. Then, the~two-point correlation function of the complex M-S variable computed on the Bunch--Davies vacuum state corresponds to ({see} \cite{bib:martin-vennin-peter-2012}, we are here dropping the prefix in $^{BD} \chi_{\mathbf{k}}^{(0)}$ for readability):
\begin{adjustwidth}{-\extralength}{0cm}
\begin{equation}
        \begin{split}
            \Xi(\mathbf{r}) &= \langle 0| v(\eta, \mathbf{x}) v(\eta, \mathbf{x+r})| 0 \rangle =\int \prod_{\mathbf{k}} d v_{\mathbf{k}}^{R} d v_{\mathbf{k}}^{I} \left( \prod_{\mathbf{k}'} \chi_{\mathbf{k'}}^{(0)*}(\eta, v_{\mathbf{k'}}) \right) v(\eta, \mathbf{x}) v(\eta, \mathbf{x+r})   \left(\prod_{\mathbf{k''}} \chi_{\mathbf{k''}}^{(0)}(\eta, v_{\mathbf{k''}}) \right)\\
            &=\left(\prod_{\mathbf{l}} |N_l(\eta)|^4\right)\int \prod_{\mathbf{k}} d v_{\mathbf{k}}^R d v_{\mathbf{k}}^I \left(\prod_{\mathbf{k'}} e^{ -2\Re(\Omega_{k'})\left[{(v_{\mathbf{k'}}^R)}^2 + {(v_{\mathbf{k'}}^I)}^2\right]} \right) v(\eta, \mathbf{x}) v(\eta, \mathbf{x+r})\\
            &= \left( \prod_\mathbf{l} \frac{2\Re(\Omega_l)}{\pi}\right) \int \frac{d \mathbf{p}}{(2\pi)^{3/2}}\int\frac{d \mathbf{q}}{(2\pi)^{3/2}} e^{i\mathbf{p}\cdot\mathbf{x}}e^{i\mathbf{q}\cdot(\mathbf{x} + \mathbf{r})} \int \prod_\mathbf{k} d v_{\mathbf{k}}^R d v_{\mathbf{k}}^I \left[v_\mathbf{p} v_\mathbf{q} e^{ -2\sum_{\mathbf{k}'}\Re(\Omega_{k'}) \left((v_{\mathbf{k}'}^R)^2 + (v_{\mathbf{k}'}^I)^2\right) } \right]
        \end{split}
    \end{equation}
\end{adjustwidth}
where we are considering each Fourier mode of the vacuum state, substituting the expression \eqref{eq:BD-state-Omega} in the second equality, and~expanding both variables in Fourier modes in the third. We observe that the last integral, due to its form, vanishes for $\mathbf{p}\neq\pm\mathbf{q}$, and~the same happens for $\mathbf{p}=\mathbf{q}$, since we obtain exponents of the form $\left[{(v_{\mathbf{p}}^R)}^2 - {(v_{\mathbf{p}}^I)}^2\right]/2$, and the real and imaginary parts contribute the same amounts. Therefore, the~surviving contribution is in the case $\mathbf{p}=-\mathbf{q}$, that is,
\begin{adjustwidth}{-\extralength}{0cm}
\begin{equation}
        \begin{split}
        \Xi(\mathbf{r}) &= \left(\prod_\mathbf{l} \frac{2\Re(\Omega_l)}{\pi}\right) \int\frac{ d \mathbf{p}}{(2\pi)^{3}} e^{-i\mathbf{p} \cdot\mathbf{r}} \; 2 \int \prod_\mathbf{k} d v_{\mathbf{k}}^R d v_{\mathbf{k}}^I \left[{(v_\mathbf{p}^R)}^2\;  e^{ -2\sum_{\mathbf{k'}}\Re(\Omega_{k'}) \left({(v_{\mathbf{k'}}^R)}^2 + {(v_{\mathbf{k'}}^I)}^2\right)}\right]\\
        &\qquad\qquad= \int\frac{d\mathbf{p}}{(2\pi)^{3}} \,e^{-i\mathbf{p}\cdot\mathbf{r}} \,\frac{1}{2\Re(\Omega_p)}
        \end{split}
    \end{equation}
\end{adjustwidth}
where we remind that $\Omega_p =\Omega_p(\eta)$ as from the definition \eqref{eq:def-Omega-BDstate}.
This corresponds, from~\eqref{eq:correl-function-amplitude} and the definition \eqref{def:power-spectrum}, to~a power spectrum of the form
\begin{linenomath}    \begin{equation}\label{eq:power-spectrum-found}
        \mathcal{P}_v(k) = \frac{k^3}{4\pi^2}\frac{1}{\Re(\Omega_k)}.
    \end{equation}
\end{linenomath} 

Therefore, the~invariant power spectrum associated with the curvature perturbation $\zeta$~\eqref{def:curvature-perturbation} is given by
\begin{linenomath}
\begin{equation}
    \mathcal{P}_\zeta(k) = \frac{4\pi G}{\epsilon \,a_0^2\, e^{2\alpha}}\mathcal{P}_v(k) = \frac{G}{\pi\epsilon}\frac{k^3}{a_0^2\, e^{2\alpha}}\frac{1}{\Re(\Omega_k)}\,.
\end{equation}
\end{linenomath}

We now evaluate this quantity in the super-Hubble limit, which in conformal time corresponds to modes for which $k\eta \to 0^-$. In~this case, we note from the definition \eqref{eq:def-Omega-BDstate} that the function $\Re(\Omega_k)$ becomes
\begin{linenomath}
\begin{equation}\label{eq:superHubblelimit-Omega}
        \Re(\Omega_k(\eta)) \approx k^3\eta^2 
    \end{equation}
\end{linenomath}
(we are using $\hbar=1$ for easier comparison with the literature).
When implementing this limit and substituting the classical solution $\alpha(\eta)$, we arrive at the following result for the primordial power spectrum in the de Sitter phase:
\begin{linenomath}
\begin{equation}
    \mathcal{P}_\zeta(k) = \frac{G\,\Hubble_\Lambda^2}{\pi \epsilon} \Bigg|_{k=a\Hubble_{\Lambda}}\,,
\end{equation}
\end{linenomath}
where $\Hubble_\Lambda = \sqrt{8\pi G \Lambda/3}$ and the slow-roll parameter $\epsilon$ is evaluated at the horizon crossing.
{Recent satellite missions, such as WMAP} \cite{bib:wmap-2013} {and PLANCK} \cite{bib:planck-results-2016,bib:planck-results-2018}, {provided an accurate detection of the fluctuation spectrum in the cosmic microwave background temperature. These observations, and in~particular, the~Gaussian profile of the fluctuations, properly fulfill the prediction of the inflation paradigm, and in~this respect, a~significant constraint for the spectral index $n_s$}
\begin{linenomath}
\begin{equation}\label{def:spectral-index}
        n_s -1 := \frac{d \ln{\mathcal{P}_{\zeta}} }{d \ln{k}}
    \end{equation}
\end{linenomath}
{is now available} \cite{bib:planck-results-2016}{. Nonetheless, some recent data analyses suggest the possibility of some anomaly in the Gaussianity of the fluctuations} \cite{bib:cabass-divalentino-melchiorri-2016} {and called attention to the possibility to be interpreted via a multifield inflationary scenario} \cite{bib:carsten-2016}.

{Clearly, the~quantum gravity corrections we are searching for are extremely small with respect to the accuracy of the current fluctuation measurements, since they are in the order of the square ratio of the inflationary energy scale to the corresponding Planckian one, namely, about $10^{-8}$. Despite the possibility of detecting such quantum gravity modifications of the spectrum in current or near-future experiments appearing unlikely, nonetheless, their prediction looks to be a fundamental conceptual challenge.}

We recall that we have here recovered the standard QFT spectrum for the primordial fluctuations via a functional approach, implementing the Gaussian fluid as a time parameter~\eqref{eq:fluid-def-time}. It is evident that the quantum gravity corrections in \eqref{eq:applic-chi1-coneta} do not modify, but~preserve the inflationary power spectrum up to this expansion order{; an analogous result derived in a different context is present in} \cite{bib:nilsson-2022}.
Such result is clearly to be attributed to the form of the modified Schr\"odinger equation \eqref{eq:applic-chi1-conS0}, which presents no coupling between the quantum gravitational degree of freedom $\alpha$ and the perturbation variables $v_{\mathbf{k}}$. It then follows that the correction to the "fast" wave function $\chi$ \eqref{eq:chi1-weight-p-alfa} factorizes, and due its time-dependent form, does not influence the evolution of the perturbation modes in the considered~setting.

\section{Towards the General~Case}\label{sec:applic-gen}
The result presented in Section~\ref{sec:powerspectrum} suggests that the quantum gravity-induced corrections on the matter evolution, obtained in the WKB expansion and via the time parameter introduced in \eqref{eq:fluid-def-time}, give as a net effect a time-dependent factor. Such term could be considered {{a posteriori}
} a phase rescaling acting on the matter wave function, as~we show here in the general~case. 

Let us start from the modified dynamics \eqref{eq:final-fluid} analyzed for a generic minisuperspace model (the supermomentum is identically vanishing); for this purpose we work with the (homogeneous) generalized variable $h_a$ (i.e., the degrees of freedom associated with the 3-geometries) and the corresponding minisupermetric $G_{ab}$, instead of the spatial metric $h_{ij}$~\cite{bib:vilenkin-1989}. We adopt for convenience the synchronous time $N=1$ such that the definition \eqref{eq:fluid-def-time} coincides with the derivative with respect to $T$, up~to a fiducial volume set to unit, but~the result here discussed stands for a generic lapse function $N$. Explicitly, the~dynamics up to the order $M^{-1}$ are described by
\begin{linenomath}
\begin{equation}\label{eq:dinamica-modif-materia}
    i\hbar \dT{\chi} = \hat{\mathcal{H}}^m \chi -i\hbar\, G_{ab} \dhij{S_0}{a} \dhij{}{b} \chi\,,
\end{equation}
\end{linenomath}
and we write the matter wave functional as
\begin{linenomath}
\begin{equation}\label{eq:separ-chi-materia}
    \chi (h_a, T, \phi) = \xi_g(h_a)\; \Theta_m (T, \phi)\,.
\end{equation}
\end{linenomath}

We remark that this is a stronger requirement and is inherently different from the Born--Oppenheimer separation \eqref{eq:psiinizialeWKB}, since $\Theta_m$ is now assumed to be independent of the generalized coordinate $h_a$. Such separation is backed by the observation that, since there is no quantum matter back-reaction in the present model, we can consider the two sets of degrees of freedom as independent. By substituting \eqref{eq:separ-chi-materia} into \eqref{eq:dinamica-modif-materia}, and~dividing by the non-trivial functional $\xi_g$, we obtain
\begin{linenomath}
\begin{equation}
    i\hbar \dT{\Theta_m} = \hat{\mathcal{H}}^m \Theta_m - \frac{i\hbar}{\xi_g}  G_{ab} \dhij{S_0}{a} \dhij{\xi_g}{b}\,  \Theta_m \,.
\end{equation}
\end{linenomath}

Here, $S_0$ belongs to the classical solution (see Equation~\eqref{eq:hamilton-jacobi}); thus, the corresponding factor is a function of time only: $\partial_{h_{a}}S_0 = f(T)$, where the form of $f$ depends on the specific cosmological model. Additionally, the~modified dynamics cannot induce dependence of $\Theta$ on the $h_{a}$, since that was separated in \eqref{eq:separ-chi-materia}. 
Then, we can express the factor containing $\xi_g$ as a constant, whose value can depend on the quantum number associated with $h_{a}$; i.e.,~its value is fixed during the dynamics once a specific foliation is selected:
\begin{linenomath}
\begin{equation}\label{eq:k-wave-number}
    \frac{1}{\xi_g} \dhij{\xi_g}{a} = i k_{(h_a)}
\end{equation}
\end{linenomath}
where for~convenience, we have inverted the couple of indices $a$ and $b$ in \eqref{eq:dinamica-modif-materia}, making use of the symmetry of the minisupermetric $G_{ab}$. The~writing $k_{(h_a)}$ is to be understood as a function of the gravitational variable $h_a$. The~solution to \eqref{eq:k-wave-number} has a plane wave structure 
\begin{linenomath}
\begin{equation}\label{eq:xi-g-plane-wave}
    \xi_g (h_a) = e^{i k_{(h_a)}\cdot  h_{a} }\,.
\end{equation}
\end{linenomath}

The functions \eqref{eq:xi-g-plane-wave} constitute a complete basis that can be adopted to construct wave packets, which will describe the quantum gravitational contribution to $\chi$. 
In what follows, we limit our attention to the plane wave \eqref{eq:xi-g-plane-wave} associated with a specific value $k_{(h_a)}$; in this case, the modified dynamics take the form
\begin{linenomath}
\begin{equation}\label{eq:modif-dyn-f-k}
    i\hbar \dT{\Theta_m} = \hat{\mathcal{H}}^m \Theta_m + \hbar f(T) \,k_{(h_a)} \Theta_m
\end{equation}
\end{linenomath}

We now rewrite the function $\Theta_m$, which is useful for the computation of the corrective effects\add{, as}:
\begin{linenomath}
\begin{equation}\label{eq:separ-theta-rho}
    \Theta_m (T,\phi) = e^{i \Lambda (T)} \varrho (T, \phi)\,,
\end{equation}
\end{linenomath}
where $\varrho$ has the same degrees of freedom with respect to $\Theta_m$, and a (complex) time-dependent phase $\Lambda$ has been separated. In~the general case, such a phase can acquire different forms depending on the wave number $k_{(h_a)}$ present in \eqref{eq:k-wave-number} and  \eqref{eq:xi-g-plane-wave} (or, as~we will discuss later, depending on the considered wave packet). It is exactly the phase factor $\Lambda(T)$ that will account for the quantum gravity corrections, since we will see that $\varrho$ exactly solves the unperturbed matter dynamics at such order. Indeed, by~substituting \eqref{eq:separ-theta-rho} into \eqref{eq:modif-dyn-f-k} and requiring that
\begin{linenomath}
\begin{equation}\label{eq:eq-for-lambda}
    \dT{\Lambda} = f(T)\, k_{(h_a)}\,,
\end{equation}
\end{linenomath}
the additional contribution on the right-hand side of \eqref{eq:modif-dyn-f-k} cancels out via the phase rescaling, and the function $\varrho$ satisfies the unperturbed Schr\"odinger evolution:
\begin{linenomath}
\begin{equation}
    i\hbar \dT{\varrho(T, \phi)} = \hat{\mathcal{H}}_m\, \varrho(T,\phi)\,.
\end{equation}
\end{linenomath}

Here, the~matter Hamiltonian $\mathcal{H}_m$ is left as a generic expression; for the purpose of the cosmological implementation above, it took the form of a time-dependent harmonic oscillator in Section~\ref{ssec:applic-alpha}.

It is then possible to discuss any effects of such quantum gravity contributions to the scalar field's power spectrum. As~previously stated, the~net effect is encased in the time-dependent phase $\Lambda(T)$ solution of \eqref{eq:eq-for-lambda}, which is actually real-valued, since $f(T)$ follows from the classical solution $S_0$. The~complete matter wave function at $\ord{M^{-1}}$ thus reads
\begin{linenomath}
\begin{equation}\label{eq:soluz-chi-matter-gen}
    \chi(h_a, T, \phi) = e^{i k_{(h_a)}\cdot  h_{a} } \, e^{i k_{(h_a)} \int dT' f(T') } \,\varrho(T,\phi)
\end{equation}
\end{linenomath}
where the integral in the second term $\int dT' f(T')$ is intended to be between values $T_0$ and $T$, for which the WKB approximation holds. We observe that the solution \eqref{eq:soluz-chi-matter-gen} has the same shape of the result 
discussed in Section~\ref{ssec:applic-alpha}. Due to the peculiar morphology of the quantum gravity factors, arising from \eqref{eq:k-wave-number} and \eqref{eq:eq-for-lambda} (which originally stem from the requirement \eqref{eq:separ-theta-rho}), the~effect on the matter spectrum is canceled once the matter wave function is properly normalized. This is the reason for which, as~shown in Section~\ref{ssec:applic-alpha}, the~quantum gravity corrections preserve the primordial inflationary~spectrum.

Clearly, the fact that, at~the order $M^{-1}$, no corrections emerge for the inflationary spectrum from quantum gravity effects, does not mean that a possible deformation of the scale invariance property cannot come out at the next orders of approximation. 
However, here is a peculiar point that deserves specific attention: the absence of a spectral modification is a consequence of the phase form that the quantum gravity corrections take in the matter wave function, and in~turn, this feature is induced by the possibility of factorizing such a wave function into a gravitational and a matter component. 
The physical meaning of this assumption must be searched in the absence of a quantum matter backreaction on the classical gravitational~background. 

\subsection{On the Role of the Matter~Backreaction}\label{ssec:backreaction}

We observe that the $S_0$ solution for the gravitational field, and in~particular, the~classical momentum term appearing in the quantum gravity corrections in Equation~\eqref{eq:dinamica-modif-materia}, do not depend, by~the considered WKB perturbation scheme, on~the quantum matter degrees of freedom. It is exactly this point which enters the possibility of factorizing the matter wave function into two independent components \eqref{eq:separ-chi-materia}. 
On the contrary, if~the H-J equation, Equation \eqref{eq:hamilton-jacobi} , contains the expectation value of the quantum matter Hamiltonian, then also the classical momentum would be, on~average, affected by the quantum degrees of freedom. Then, the~choice of a factorized form for the matter wave function, even if still possible, would no longer appear as a natural solution to the perturbed~dynamics.

To elucidate this point of view, we here discuss in more detail the role played by the matter backreaction. In~fact, when implementing a standard B-O scheme (see the original formulation~\cite{bib:born-1927}) in the WKB approximation order by order in $1/M$ \cite{bib:bertoni-venturi-1996}, it is immediately recognizable that the quantum matter expectation value enters both the right-hand side of the H-J equation\remove{, Equation} \eqref{eq:hamilton-jacobi}, and the Schr\"odinger equation\remove{, Equation} \eqref{eq:fluid-schrod-eq-M0} (see also~\cite{bib:massar-1998} and for a review~\cite{bib:schander-thiemann-2021}). 
As stated in~\cite{bib:kiefer-QG}, this contribution can be 
easily removed from the Schr\"odinger dynamics by phase rescaling, where the phase contains the matter backreaction term, though as a function of the gravitational degrees of freedom only. This operation is allowed by a natural gauge invariance of the total B-O wave function. 
However, as~shown in~\cite{bib:montani-digioia-maniccia-2021}, this redefinition of the matter wave function induces an opposite change of phase to the gravitational one, with~the net effect that the backreaction term is also removed from the H-J~equation. 

These considerations suggest that such a contribution could always be neglected in view of the gauge invariance analyzed above, and hence, that the emergence of a quantum gravity correction originating from the matter backreaction cannot be inferred.
Nonetheless, we question here the correctness of doing such a phase redefinition via the matter expectation value. 
Actually, in~the B-O procedure, the~gauge invariance is used to eliminate the Berry phase~\cite{bib:mead-1992-geomphase,bib:panati-2007,bib:seung-PRL-2014}, but~not to cancel the (fast) electronic eigenvalue contribution in the  (slow) nuclear dynamics~\cite{bib:bransden}.
From this point of view, it is more natural to maintain the expectation value contribution both in the H-J equation and in the Schr\"odinger one. This would lead to a non-trivial coupled integro-partial differential system which could be treated with a self-consistent method (for a related treatment of the backreaction in a different context, see~\cite{bib:cheng-2022}). 

The discussion above was thought to refer to order $M^1$ and $M^0$ of the WKB approximation, but~it naturally extends to the $M^{-1}$ order. 
Thus, if~we include the matter Hamiltonian term in Equation~\eqref{eq:dinamica-modif-materia}, we arrive to a coupled system that only at the lowest order of approximation in a Hartree self-consistent approach can be reduced to the form discussed in Section~\ref{ssec:fluidWKB}. The~complete problem naturally introduces dependence of the H-J function $S_0$ on the matter one (via an integral of the matter's degrees of freedom); 
 this point clarifies the technical content of the discussion above on the role of the matter backreaction in the separability of  Equation \eqref{eq:dinamica-modif-materia} to some order of approximation.
Therefore, we are led to conclude that the proposed WKB expansion in the quantity $1/M$ must carefully take into account the evaluation of the matter (average) backreaction on the gravitational quasi-classical~background.

\section{Concluding~Remarks}\label{sec:conclusions}

Here, we reviewed the analysis presented in~\cite{bib:maniccia-montani-2022}, aiming to calculate the quantum gravity corrections to QFT, in~the theoretical framework of fixing a Gaussian reference frame, as discussed in~\cite{bib:kuchar-torre-1991}. 
The motivations for addressing such a revised scheme came from the search for a formulation which is not affected by the non-unitarity questions faced in~\cite{bib:kiefer-1991,bib:kiefer-2018} when reconstructing a time variable for the matter wave function from the classical limit of the background gravitational field.
The physical clock in~\cite{bib:maniccia-montani-2022} is provided by the materialization of the Gaussian frame as a dust fluid. It is important to recall here that such an emerging dust-like contribution no longer has, in~the WKB expansion, the~shortcoming of a non-positive defined energy~density. 

The present study implemented the procedure mentioned above to calculate the possible quantum gravity corrections to the primordial inflationary spectrum. 
We considered the quasi-classical background corresponding to a Robertson--Walker geometry in the presence of a cosmological constant term, mimicking the vacuum energy of an inflationary phase transition, as~viewed in the resulting de Sitter evolution. 
The matter field we quantized in the proposed scheme was clearly the inflaton scalar degree of freedom, for~which we applied a Fourier decomposition and introduced the gauge-invariant M-S formulation~\mbox{\cite{bib:mukhanov-1985,bib:sasaki-1986,bib:mukhanov-1988}.} 
The field would have been in principle associated with a wave functional, describing its dynamics, but~the Fourier decomposition of its Hamiltonian allowed us to deal with a minisuperspace formulation for each independent wavenumber modulus (we recall that the inflaton can be regarded, with~a very good approximation, as~a free massless scalar field during the slow-rolling phase of the inflation process). 
Clearly, we solved the wave equation amended for the quantum gravity corrections, and we ended up showing that the solution of the time-dependent harmonic oscillator associated with each $k$-mode is rescaled by a phase factor, due to the additional contribution to the Schr\"odinger equation. 
As an immediate consequence, we could conclude that, at~the considered approximation in the WKB scheme, no modification of the inflationary spectrum of the Universe can be~determined. 

{It would be worth analyzing the present formulation in the case in which the background gravitational field is described by a modified theory of gravity. For~a discussion of how modified gravity affects the inflationary spectrum, see} \cite{bib:hwang-noh-1996,bib:hwang-1997,bib:hwang-noh-2001,bib:hwang-noh-2005,bib:capozziello-2008,bib:capozziello-montani-2011}{\change{. We must consider further investigation of how}{but it calls attention for further investigation how} these results would appear in the present framework, i.e.,~including quantum gravity corrections of the extended formulation (for approaches which quantize the modified metric $f(R)$ gravity, see} \cite{bib:bamonti-costantini-montani-2022,bib:deangelis-figurato-montani-2021,bib:montani-deangelis-2022}{).}

We then discussed this result in the scheme of a generic minisuperspace model, in~order to outline the real physical explanation for such surprising preservation of the scale-invariant spectrum in the theory proposed in~\cite{bib:maniccia-montani-2022}. 
From a mathematical point of view, we recognized that the emergence of a phase term in the matter wave function, depending on the scale factor, is a consequence of the possibility of factorizing such a wave function ab initio. 
Clearly, by considering higher-order contributions to the inflaton Schr\"odinger equation in the expansion with respect to $1/M$, a~modification of the scale invariant spectrum could arise. 
However, in~the considered theoretical framework, the~factorization of the matter wave function came from the absence of an average matter backreaction in the classical Robertson--Walker dynamics, 
i.e.,~in the H-J equation, as~discussed in detail in Section~\ref{ssec:backreaction}. {A study of the modifications to the power spectrum when the backreaction is taken into account is beyond the analysis here presented and could be investigated in future works.}

The analysis above suggests that the scheme in~\cite{bib:maniccia-montani-2022} could require further restatement in order to better separate the classical and quantum degrees of freedom{, which is beyond the scope of this paper}. 
Particular attention has to be focused on the procedure by which the gravitational degrees of freedom are treated---i.e.,~their classical and quantum components would have to be described via independent variables{;} see, for example, the proposal \cite{bib:maniccia-montani-antonini-2023}. Only after such a reformulation of the gravitational background could the~question concerning the matter backreaction be properly addressed in the B-O WKB picture  proposed here. This perspective should call attention for future developments calculating the quantum gravity corrections to the inflationary~spectrum.

\authorcontributions{{All} 
 the authors provided equivalent contributions to the scientific content and editing of the~manuscript. All authors have read and agreed to the published version of the manuscript.}

\funding{This research received no external~funding.}

\dataavailability{Not applicable.}

\acknowledgments{G. Maniccia thanks the TAsP INFN initiative for~support.}

\conflictsofinterest{The authors declare no conflict of~interest.} 

\appendixtitles{yes}
\appendixstart
\appendix
\section{The Lewis--Riesenfeld Invariant~Method}\label{appendix:Lewis}

The so-called Lewis--Riesenfeld invariant method~\cite{bib:lewis-1967,bib:lewis-1968,bib:lewis-riesenfeld-1969} represents an algorithm for computing the solution for a time-dependent quantum system, in~the cases in which a specific invariant can be identified. Generally speaking, given a system with a generic time-dependent Hamiltonian $\mathcal{H}(t)$, the~determination of a Hermitian invariant $I$ (also called Lewis--Riesenfeld invariant) associated with $\hat{\mathcal{H}}(t)$ gives an eigenstate basis that can be used to obtain the solution's wave function. Here, we show the application of this method for the time-dependent quantum harmonic oscillator, for~which the method was first~developed.

Starting from the time-dependent harmonic Hamiltonian~\eqref{eq:TDharmonic-oscillator}, one can check that the invariant corresponds to the following expression:
\begin{linenomath}
\begin{equation}\label{eq:invariant-I}
        I = \frac{1}{2}\left[\frac{v_{\mathbf{k}}^2}{\rho_k^2} + (\rho_k \pi_{v_{\mathbf{k}}} - \dot{\rho}_k v_{\mathbf{k}})^2 \right]
    \end{equation}
\end{linenomath}
where $\rho_k$ satisfies the so-called Ermakov equation:
\begin{linenomath}
\begin{equation}
        \ddot{\rho}_k + \omega_k^2 \rho_k = \frac{1}{\rho_k^3}
    \end{equation}
\end{linenomath}
and we recall that the time-dependence is inside $\omega_k(\eta)$, as~is the case in Section~\ref{ssec:applic-alpha} (see the definition \eqref{eq:def-omega-TDHA}). The~solution for $\rho_k$ is explicitly 
\begin{linenomath}
\begin{equation}\label{eq:rho-solution}
        \begin{split}
        \rho_k = \gamma_1 &\left[A^2\frac{(\eta  k \sin (\eta  k)+\cos (\eta  k))^2}{\eta ^2 k^3}+B^2 \frac{(\eta  k \cos (\eta  k)-\sin (\eta  k))^2}{\eta ^2 k^3} \right.\\
        &\left.+ \gamma _2 \sqrt{A^2 B^2-1}\; \frac{(\eta k \sin(\eta k)+\cos(\eta k))(\eta k \cos(\eta k)-\sin(\eta k))}{\eta^2 k^3}\right]^{\frac{1}{2}}
        \end{split}
    \end{equation}
\end{linenomath}
where $A$, $B$, and $\gamma_1=\gamma_2 = \pm 1$ are constants to be appropriately chosen in the cosmological scenario. 
The expression \eqref{eq:rho-solution} will allow one to find the eigenstates of \eqref{eq:invariant-I}, which will be described, for~each mode, by~a quantum index $n$:
\begin{linenomath}
\begin{equation}\label{eq:invariant-I-basis}
    \hat{I}\, \phi_{n,\mathbf{k}}(\eta,v_{\mathbf{k}}) = \lambda_n \,\phi_{n,\mathbf{k}}(\eta,v_{\mathbf{k}})\,.
    \end{equation}
\end{linenomath}

The eigenstates can be determined by applying the following unitary transformation:
\begin{linenomath}
\begin{equation}\label{eq:phi-tilde-tilde}
    \exp\left({-\frac{i}{2\hbar}\frac{\Dot{\rho}_k}{\rho_k}}v_{\mathbf{k}}^2\right)\phi_{n,\mathbf{k}}  = \frac{1}{\rho_k^{1/2}}\Tilde{\phi}_{n,\mathbf{k}} \,,
    \end{equation}
\end{linenomath}
that transforms Equation~\eqref{eq:TDharmonic-oscillator} into 
\begin{linenomath}
\begin{equation}
    \left(-\frac{\hbar}{2}\partial_{\mathrm{v}_{\mathbf{k}}}^2 + \frac{\mathrm{v}_{\mathbf{k}}}{2}\right)\Tilde{\phi}_{n,\mathbf{k}} = \lambda_n\Tilde{\phi}_{n,\mathbf{k}}
    \end{equation}
\end{linenomath}
where $\mathrm{v}_{\mathbf{k}} = v_{\mathbf{k}}/\rho_k$. Such an equation is easily solved: the eigenvalues are of the form
\begin{linenomath}
\begin{equation}
        \lambda_n = \hbar\left(n+\frac{1}{2}\right)\,,
    \end{equation}
\end{linenomath}
coinciding with the eigenvalues of the invariant $I$ (see \eqref{eq:invariant-I-basis}), without~an explicit dependence on $\mathbf{k}$, and the corresponding eigenstates $\Tilde{\phi}_{n,\mathbf{k}}$ are rescaled back from \eqref{eq:phi-tilde-tilde} to give the invariant eigenstates
\begin{adjustwidth}{-\extralength}{0cm}
\begin{equation}\label{eq:sol-invariant-eigenstates}
        \phi_{n,\mathbf{k}}(\eta,v_{\mathbf{k}}) =\left[\frac{1}{(\pi\hbar)^{1/2} 2^n n!\,\rho_k(\eta)}\right]^{1/2} \exp \left[\frac{i}{2\hbar}\left(\frac{\dot{\rho}_k(\eta)}{\rho_k(\eta)}+\frac{i}{ \rho_k^2(\eta)}\right) v_{\mathbf{k}}^2\right] H_n\left(\frac{1}{\hbar^{1/2}}\frac{v_{\mathbf{k}}}{\rho_k(t)}\right)\,.
    \end{equation}
\end{adjustwidth}

Here, $H_n$ are the Hermite polynomials. The~state basis \eqref{eq:sol-invariant-eigenstates} allows one to write the solution for the starting time-dependent harmonic oscillator 
\eqref{eq:TDharmonic-oscillator} as
\begin{linenomath}
    \begin{gather}
    \chi^{(0)}_{\mathbf{k}} (\eta,v_{\mathbf{k}}) = \sum_n c_{n,k} e^{i\delta_{n,k}(\eta)}\phi_{n,\mathbf{k}}(\eta,v_{\mathbf{k}}), \label{eq:basis-chi0-invariant-eigenstates}\\
    \delta_{n,k}(\eta) =-\left(n+\frac{1}{2}\right)\int d\eta \,\frac{1}{\rho_k^2(\eta)} \,,\label{eq:delta-nk-invariant}
    \end{gather}
\end{linenomath}
where $c_{n,k}$ are some suitable coefficients fixed by the system's boundary conditions. Equation~\eqref{eq:basis-chi0-invariant-eigenstates} is thus the wave function describing the evolution of the time-dependent harmonic oscillator system.
\begin{adjustwidth}{-\extralength}{0cm}

\reftitle{References}


\end{adjustwidth}

\begin{thebibliography}{999}

\bibitem[Thiemann(2007)]{bib:thiemann-book}
Thiemann, T.
\newblock {\em Modern Canonical Quantum General Relativity}; Cambridge
  Monographs on Mathematical Physics, {Cambridge University Press: Cambridge, England,} 
  2007.
\newblock {\url{https://doi.org/10.1017/CBO9780511755682}}.

\bibitem[Cianfrani et~al.(2014)Cianfrani, Lecian, Lulli, and
  Montani]{bib:cianfrani-canonicalQuantumGravity}
Cianfrani, F.; Lecian, O.M.; Lulli, M.; Montani, G.
\newblock {\em Canonical Quantum Gravity}; {World Scientific: Singapore,}  2014. 
\newblock {\url{https://doi.org/10.1142/8957}}.

\bibitem[{DeWitt}(1967)]{bib:dewitt1-1967}
{DeWitt}, B.S.
\newblock Quantum Theory of Gravity. {I}. The Canonical Theory.
\newblock {\em Phys.\ Rev.} {\bf 1967}, {\em 160},~1113--1148.
\newblock {\url{https://doi.org/10.1103/PhysRev.160.1113}}.

\bibitem[DeWitt(1967{\natexlab{a}})]{bib:dewitt2-1967}
DeWitt, B.S.
\newblock Quantum Theory of Gravity. {II}. The Manifestly Covariant Theory.
\newblock {\em Phys. Rev.} {\bf 1967}, {\em 162},~1195--1239.
\newblock {\url{https://doi.org/10.1103/PhysRev.162.1195}}.

\bibitem[DeWitt(1967{\natexlab{b}})]{bib:dewitt3-1967}
DeWitt, B.S.
\newblock Quantum Theory of Gravity. {III}. Applications of the Covariant
  Theory.
\newblock {\em Phys. Rev.} {\bf 1967}, {\em 162},~1239--1256.
\newblock {\url{https://doi.org/10.1103/PhysRev.162.1239}}.

\bibitem[Kucha\u{r}(1980)]{bib:kuchar-1981}
Kucha\u{r}, K.V.
\newblock Canonical Methods of Quantization.
\newblock In Proceedings of the Oxford Conference on Quantum Gravity,  Oxford, UK, 15--19 April 1980; pp. 329--376.

\bibitem[Isham(1993)]{bib:isham-book-1993}
Isham, C.J. Canonical Quantum Gravity and the Problem of Time.
\newblock In {\em Integrable Systems, Quantum Groups, and Quantum Field
  Theories}; Springer: Dordrecht, The Netherlands,  1993; pp. 157--287.
\newblock {\url{https://doi.org/10.1007/978-94-011-1980-1_6}}.

\bibitem[Wald(1993)]{bib:wald-1993}
Wald, R.M.
\newblock Proposal for solving the ``problem of time'' in canonical quantum
  gravity.
\newblock {\em Phys. Rev. D} {\bf 1993}, {\em 48},~R2377--R2381.
\newblock {\url{https://doi.org/10.1103/physrevd.48.r2377}}.

\bibitem[{Rovelli}(1991)]{bib:rovelli-1991-time}
{Rovelli}, C.
\newblock Time in quantum gravity: An hypothesis.
\newblock {\em Phys.\ Rev.\ D} {\bf 1991}, {\em 43},~442--456.
\newblock {\url{https://doi.org/10.1103/PhysRevD.43.442}}.

\bibitem[Kucha\v{r}(2011)]{bib:kuchar2011-review}
Kucha\v{r}, K.V.
\newblock Time and interpretations of Quantum Gravity.
\newblock {\em Int. J. Mod. Phys. D} {\bf 2011}, {\em
  20},~3--86.
\newblock {\url{https://doi.org/10.1142/S0218271811019347}}.

\bibitem[Ashtekar(1986)]{bib:ashtekar-1986}
Ashtekar, A.
\newblock New Variables for Class. Quantum Gravity.
\newblock {\em Phys. Rev. Lett.} {\bf 1986}, {\em 57},~2244--2247.
\newblock {\url{https://doi.org/10.1103/PhysRevLett.57.2244}}.

\bibitem[Rovelli(2004)]{bib:rovelli-book-QG}
Rovelli, C.
\newblock {\em Quantum Gravity}; Cambridge Monographs on Mathematical Physics,
  {Cambridge University Press: Cambridge, England}  2004.
\newblock {\url{https://doi.org/10.1017/CBO9780511755804}}.

\bibitem[Rovelli and Smolin(1995)]{bib:rovelli-smolin-1995}
Rovelli, C.; Smolin, L.
\newblock Spin networks and quantum gravity.
\newblock {\em Phys. Rev. D} {\bf 1995}, {\em 52},~5743--5759.
\newblock {\url{https://doi.org/10.1103/PhysRevD.52.5743}}.

\bibitem[Nicolai et~al.(2005)Nicolai, Peeters, and
  Zamaklar]{bib:nicolai-review-2005}
Nicolai, H.; Peeters, K.; Zamaklar, M.
\newblock Loop quantum gravity: An outside view.
\newblock {\em Class. Quantum Gravity} {\bf 2005}, {\em 22},~R193.
\newblock {\url{https://doi.org/10.1088/0264-9381/22/19/R01}}.

\bibitem[Birrell and Davies(1982)]{bib:birrel-davies}
Birrell, N.D.; Davies, P.C.W.
\newblock {\em {Quantum Fields in Curved Space}}; {Cambridge Monographs on
  Mathematical Physics}, {Cambridge University Press: Cambridge, England}  1982.
\newblock {\url{https://doi.org/10.1017/CBO9780511622632}}.

\bibitem[Wald(1995{\natexlab{a}})]{bib:book-wald-QFT-curved-space}
Wald, R.M.
\newblock {\em {Quantum Field Theory in Curved Space-Time and Black Hole
  Thermodynamics}}; {Chicago Lectures in Physics}, University of Chicago Press:
  Chicago, IL, USA, 1995.

\bibitem[Wald(1995{\natexlab{b}})]{bib:wald-review-1995}
Wald, R.M.
\newblock {Quantum Field Theory in Curved Spacetime} \emph{arXiv}  \textbf{1995},  	arXiv:gr-qc/9509057.
\newblock {\url{https://doi.org/10.48550/ARXIV.GR-QC/9509057}}.

\bibitem[Crispino et~al.(2008)Crispino, Higuchi, and
  Matsas]{bib:crispino-review-unruh-2008}
Crispino, L.C.B.; Higuchi, A.; Matsas, G.E.A.
\newblock {The Unruh effect and its applications}.
\newblock {\em Rev. Mod. Phys.} {\bf 2008}, {\em 80},~787--838.
\newblock {\url{https://doi.org/10.1103/revmodphys.80.787}}.

\bibitem[Hawking(1975)]{bib:hawking-1975}
Hawking, S.W.
\newblock {Particle Creation by Black Holes}.
\newblock {\em Commun. Math. Phys.} {\bf 1975}, {\em 43},~199--220.
\newblock Erratum in  \emph{Commun. Math. Phys}. \textbf{1976}, \emph{46}, 206.
  {\url{https://doi.org/10.1007/BF02345020}}.

\bibitem[{Vilenkin}(1989)]{bib:vilenkin-1989}
{Vilenkin}, A.
\newblock Interpretation of the wave function of the Universe.
\newblock {\em Phys.\ Rev.\ D} {\bf 1989}, {\em 39},~1116--1122.
\newblock {\url{https://doi.org/10.1103/PhysRevD.39.1116}}.

\bibitem[{Kiefer} and {Singh}(1991)]{bib:kiefer-1991}
{Kiefer}, C.; {Singh}, T.P.
\newblock Quantum gravitational corrections to the functional {Schr\"odinger}
  equation.
\newblock {\em Phys.\ Rev.\ D} {\bf 1991}, {\em 44},~1067--1076.
\newblock {\url{https://doi.org/10.1103/PhysRevD.44.1067}}.

\bibitem[Barvinsky(1993)]{bib:barvinsky-1993}
Barvinsky, A.
\newblock Unitarity approach to quantum cosmology.
\newblock {\em Phys. Rep.} {\bf 1993}, {\em 230},~237--367.
\newblock {\url{https://doi.org/10.1016/0370-1573(93)90032-9}}.

\bibitem[Vilenkin(1994)]{bib:vilenkin-1994-rassegna}
Vilenkin, A.
\newblock Approaches to quantum cosmology.
\newblock {\em Phys. Rev. D} {\bf 1994}, {\em 50},~2581--2594.
\newblock {\url{https://doi.org/10.1103/PhysRevD.50.2581}}.

\bibitem[{Ohkuwa}(1995)]{bib:okhuwa-1995}
{Ohkuwa}, Y.
\newblock {Time in the semi-classical approximation to quantum cosmology.}
\newblock {\em Nuovo C. B Ser.} {\bf 1995}, {\em 110B},~53--60.
\newblock {\url{https://doi.org/10.1007/BF02741289}}.

\bibitem[{Bertoni} et~al.(1996){Bertoni}, {Finelli}, and
  {Venturi}]{bib:bertoni-venturi-1996}
{Bertoni}, C.; {Finelli}, F.; {Venturi}, G.
\newblock The {Born-Oppenheimer} approach to the matter-gravity system and
  unitarity.
\newblock {\em Class. Quantum Gravity} {\bf 1996}, {\em 13},~2375--2383.
\newblock {\url{https://doi.org/10.1088/0264-9381/13/9/005}}.

\bibitem[Brizuela et~al.(2016)Brizuela, Kiefer, and
  Kr\"amer]{bib:brizuela-kiefer-2016-desitter}
Brizuela, D.; Kiefer, C.; Kr\"amer, M.
\newblock Quantum-gravitational effects on gauge-invariant scalar and tensor
  perturbations during inflation: The de Sitter case.
\newblock {\em Phys. Rev. D} {\bf 2016}, {\em 93},~104035.
\newblock {\url{https://doi.org/10.1103/PhysRevD.93.104035}}.

\bibitem[{Brizuela} et~al.(2016){Brizuela}, {Kiefer}, and
  {Krämer}]{bib:brizuela-kiefer-2016-slow-roll}
{Brizuela}, D.; {Kiefer}, C.; {Krämer}, M.
\newblock Quantum-gravitational effects on gauge-invariant scalar and tensor
  perturbations during inflation: The slow-roll approximation.
\newblock {\em Phys.\ Rev.\ D} {\bf 2016}, {\em 94},~123527.
\newblock {\url{https://doi.org/10.1103/PhysRevD.94.123527}}.

\bibitem[Kamenshchik et~al.(2017)Kamenshchik, Tronconi, and
  Venturi]{bib:venturi-2017}
Kamenshchik, A.Y.; Tronconi, A.; Venturi, G.
\newblock The {Born–Oppenheimer} method, quantum gravity and matter.
\newblock {\em Class. Quantum Gravity} {\bf 2017}, {\em 35},~015012.
\newblock {\url{https://doi.org/10.1088/1361-6382/aa8fb3}}.

\bibitem[{Kiefer} and {Wichmann}(2018)]{bib:kiefer-2018}
{Kiefer}, C.; {Wichmann}, D.
\newblock Semiclassical approximation of the {Wheeler--DeWitt} equation:
  Arbitrary orders and the question of unitarity.
\newblock {\em Gen. Relativ. Gravit.} {\bf 2018}, {\em 50},~66.
\newblock {\url{https://doi.org/10.1007/s10714-018-2390-4}}.

\bibitem[Kamenshchik et~al.(2020)Kamenshchik, Tronconi, and
  Venturi]{bib:venturi-2020}
Kamenshchik, A.Y.; Tronconi, A.; Venturi, G.
\newblock Quantum cosmology and the inflationary spectra from a nonminimally
  coupled inflaton.
\newblock {\em Phys. Rev. D} {\bf 2020}, {\em 101},~023534.
\newblock {\url{https://doi.org/10.1103/physrevd.101.023534}}.

\bibitem[Rotondo(2020)]{bib:rotondo-2020}
Rotondo, M.
\newblock The Functional {Schrödinger} Equation in the Semiclassical Limit of
  Quantum Gravity with a Gaussian Clock Field.
\newblock {\em Universe} {\bf 2020}, {\em 6},~176.
\newblock {\url{https://doi.org/10.3390/universe6100176}}.

\bibitem[Gielen(2021)]{bib:gielen-2021}
Gielen, S.
\newblock Frozen formalism and canonical quantization in group field theory.
\newblock {\em Phys. Rev. D} {\bf 2021}, {\em 104},~106011.
\newblock {\url{https://doi.org/10.1103/physrevd.104.106011}}.

\bibitem[Di~Gioia et~al.(2021)Di~Gioia, Maniccia, Montani, and
  Niedda]{bib:montani-digioia-maniccia-2021}
Di~Gioia, F.; Maniccia, G.; Montani, G.; Niedda, J.
\newblock Nonunitarity problem in quantum gravity corrections to quantum field
  theory with Born-Oppenheimer approximation.
\newblock {\em Phys. Rev. D} {\bf 2021}, {\em 103},~103511.
\newblock {\url{https://doi.org/10.1103/PhysRevD.103.103511}}.

\bibitem[Maniccia and Montani(2023)]{bib:maniccia-montani-2021}
Maniccia, G.; Montani, G. WKB approach to the gravity-matter dynamics: A
  cosmological implementation.
\newblock In {\em The Sixteenth Marcel Grossmann Meeting}; {World Scientific: Singapore} 
  2023; pp. 4146--4158.
\newblock {\url{https://doi.org/10.1142/9789811269776_0345}}.

\bibitem[Maniccia and Montani(2022)]{bib:maniccia-montani-2022}
Maniccia, G.; Montani, G.
\newblock Quantum gravity corrections to the matter dynamics in the presence of
  a reference fluid.
\newblock {\em Phys. Rev. D} {\bf 2022}, {\em 105},~086014.
\newblock {\url{https://doi.org/10.1103/PhysRevD.105.086014}}.

\bibitem[Maniccia et~al.(2022)Maniccia, De~Angelis, and
  Montani]{bib:maniccia-deangelis-montani-review-2022}
Maniccia, G.; De~Angelis, M.; Montani, G.
\newblock WKB Approaches to Restore Time in Quantum Cosmology: Predictions and
  Shortcomings.
\newblock {\em Universe} {\bf 2022}, {\em 8},~556.
\newblock {\url{https://doi.org/10.3390/universe8110556}}.

\bibitem[Lapchinsky and Rubakov(1979)]{bib:rubakov-lapchinsky-1979}
Lapchinsky, V.G.; Rubakov, V.A.
\newblock {Canonical Quantization of Gravity and Quantum Field Theory in Curved
  Space-time}.
\newblock {\em Acta Phys. Polon. B} {\bf 1979}, {\em 10},~1041--1048.

\bibitem[Agostini et~al.(2017)Agostini, Cianfrani, and
  Montani]{bib:agostini-cianfrani-montani-2017}
Agostini, L.; Cianfrani, F.; Montani, G.
\newblock Probabilistic interpretation of the wave function for the Bianchi I
  model.
\newblock {\em Phys. Rev. D} {\bf 2017}, {\em 95},~126010.
\newblock {\url{https://doi.org/10.1103/PhysRevD.95.126010}}.

\bibitem[Banks(1985)]{bib:banks-1985}
Banks, T.
\newblock TCP, quantum gravity, the cosmological constant and all that.
\newblock {\em Nucl. Phys. B} {\bf 1985}, {\em 249},~332--360.
\newblock {\url{https://doi.org/10.1016/0550-3213(85)90020-3}}.

\bibitem[Vilenkin(2002)]{bib:vilenkin-2002}
Vilenkin, A.
\newblock {Quantum cosmology and eternal inflation}.
\newblock In Proceedings of the {Workshop on Conference on the Future of
  Theoretical Physics and Cosmology in Honor of Steven Hawking's 60th
  Birthday}, Cambridge, UK,  7--10 January  2002; pp. 649--666.
\newblock {\url{https://doi.org/10.48550/ARXIV.GR-QC/0204061}}.

\bibitem[Battisti et~al.(2009)Battisti, Belvedere, and
  Montani]{bib:battisti-belvedere-montani-2009}
Battisti, M.V.; Belvedere, R.; Montani, G.
\newblock Semiclassical suppression of weak anisotropies of a generic Universe.
\newblock {\em {EPL} (Europhys. Lett.)} {\bf 2009}, {\em 86},~69001.
\newblock {\url{https://doi.org/10.1209/0295-5075/86/69001}}.

\bibitem[Kiefer(2013)]{bib:kiefer-2013-review}
Kiefer, C.
\newblock Conceptual Problems in Quantum Gravity and Quantum Cosmology.
\newblock {\em {ISRN} Math. Phys.} {\bf 2013}, {\em 2013},~509316.
\newblock {\url{https://doi.org/10.1155/2013/509316}}.

\bibitem[Moriconi and Montani(2017)]{bib:moriconi-montani-2017}
Moriconi, R.; Montani, G.
\newblock Behavior of the Universe anisotropy in a big-bounce cosmology.
\newblock {\em Phys. Rev. D} {\bf 2017}, {\em 95},~123533.
\newblock {\url{https://doi.org/10.1103/PhysRevD.95.123533}}.

\bibitem[Montani et~al.(2018)Montani, Marchi, and
  Moriconi]{bib:montani-marchi-moriconi-2018}
Montani, G.; Marchi, A.; Moriconi, R.
\newblock Bianchi I model as a prototype for a cyclical Universe.
\newblock {\em Phys. Lett. B} {\bf 2018}, {\em 777},~191--200.
\newblock {\url{https://doi.org/10.1016/j.physletb.2017.12.016}}.

\bibitem[Damour and Vilenkin(2019)]{bib:damour-vilenkin-2019}
Damour, T.; Vilenkin, A.
\newblock Quantum instability of an oscillating universe.
\newblock {\em Phys. Rev. D} {\bf 2019}, {\em 100},~083525.
\newblock {\url{https://doi.org/10.1103/PhysRevD.100.083525}}.

\bibitem[Kiefer et~al.(2019)Kiefer, Kwidzinski, and Piontek]{bib:kiefer-2019}
Kiefer, C.; Kwidzinski, N.; Piontek, D.
\newblock Singularity avoidance in Bianchi I quantum cosmology.
\newblock {\em  Eur. Phys. J. C} {\bf 2019}, {\em 79}, 1--12.
\newblock {\url{https://doi.org/10.1140/epjc/s10052-019-7193-6}}.

\bibitem[Chiovoloni et~al.(2020)Chiovoloni, Montani, and
  Cascioli]{bib:montani-chiovoloni-cascioli-2020}
Chiovoloni, R.; Montani, G.; Cascioli, V.
\newblock Quantum dynamics of the corner of the Bianchi IX model in the WKB
  approximation.
\newblock {\em Phys. Rev. D} {\bf 2020}, {\em 102},~083519.
\newblock {\url{https://doi.org/10.1103/PhysRevD.102.083519}}.

\bibitem[De~Angelis and Montani(2020)]{bib:deangelis-2020}
De~Angelis, M.; Montani, G.
\newblock Dynamics of quantum anisotropies in a Taub universe in the WKB
  approximation.
\newblock {\em Phys. Rev. D} {\bf 2020}, {\em 101},~103532.
\newblock {\url{https://doi.org/10.1103/PhysRevD.101.103532}}.

\bibitem[Robles-Pérez(2021)]{bib:robles-perez-2021}
Robles-Pérez, S.J.
\newblock Quantum Cosmology with Third Quantisation.
\newblock {\em Universe} {\bf 2021}, {\em 7},~404.
\newblock {\url{https://doi.org/10.3390/universe7110404}}.

\bibitem[Kiefer and Peter(2022)]{bib:peter-kiefer-2022}
Kiefer, C.; Peter, P.
\newblock Time in Quantum Cosmology.
\newblock {\em Universe} {\bf 2022}, {\em 8},~36.
\newblock {\url{https://doi.org/10.3390/universe8010036}}.

\bibitem[Sahlmann and
  Thiemann(2006{\natexlab{a}})]{bib:sahlmann-thiemann-2006-1}
Sahlmann, H.; Thiemann, T.
\newblock {Towards the QFT on curved spacetime limit of QGR: I. A general
  scheme}.
\newblock {\em Class. Quantum Gravity} {\bf 2006}, {\em 23},~867.
\newblock {\url{https://doi.org/10.1088/0264-9381/23/3/019}}.

\bibitem[Sahlmann and
  Thiemann(2006{\natexlab{b}})]{bib:sahlmann-thiemann-2006-2}
Sahlmann, H.; Thiemann, T.
\newblock {Towards the QFT on curved spacetime limit of QGR: II. A concrete
  implementation}.
\newblock {\em Class. Quantum Gravity} {\bf 2006}, {\em 23},~909.
\newblock {\url{https://doi.org/10.1088/0264-9381/23/3/020}}.

\bibitem[Ashtekar et~al.(2009)Ashtekar, Kaminski, and
  Lewandowski]{bib:ashtekar-leewandowski-2009}
Ashtekar, A.; Kaminski, W.; Lewandowski, J.
\newblock Quantum field theory on a cosmological, quantum space-time.
\newblock {\em Phys. Rev. D} {\bf 2009}, {\em 79},~064030.

\bibitem[Dapor et~al.(2012)Dapor, Lewandowski, and
  Tavakoli]{bib:lewandowski-2012}
Dapor, A.; Lewandowski, J.; Tavakoli, Y.
\newblock Lorentz symmetry in QFT on quantum Bianchi I space-time.
\newblock {\em Phys. Rev. D} {\bf 2012}, {\em 86},~064013.
\newblock {\url{https://doi.org/10.1103/PhysRevD.86.064013}}.

\bibitem[Bojowald and Halnon(2018)]{bib:bojowald-2018}
Bojowald, M.; Halnon, T.
\newblock Time in quantum cosmology.
\newblock {\em Phys. Rev. D} {\bf 2018}, {\em 98},~066001.
\newblock {\url{https://doi.org/10.1103/PhysRevD.98.066001}}.

\bibitem[Chataignier and Krämer(2021)]{bib:chataignier-kramer-2021}
Chataignier, L.; Krämer, M.
\newblock Unitarity of quantum-gravitational corrections to primordial
  fluctuations in the Born-Oppenheimer approach.
\newblock {\em Phys.\ Rev.\ D} {\bf 2021}, {\em 103},~066005.
\newblock {\url{https://doi.org/10.1103/physrevd.103.066005}}.

\bibitem[Gielen and Men{\'{e}}ndez-Pidal(2022)]{bib:gielen-2022}
Gielen, S.; Men{\'{e}}ndez-Pidal, L.
\newblock Unitarity, clock dependence and quantum recollapse in quantum
  cosmology.
\newblock {\em Class. Quantum Gravity} {\bf 2022}, {\em 39},~075011.
\newblock {\url{https://doi.org/10.1088/1361-6382/ac504f}}.

\bibitem[Montani(2002)]{bib:montani-2002}
Montani, G.
\newblock Canonical quantization of gravity without “frozen formalism”.
\newblock {\em Nucl. Phys. B} {\bf 2002}, {\em 634},~370--392.
\newblock {\url{https://doi.org/10.1016/S0550-3213(02)00301-2}}.

\bibitem[{Kucha\u{r}} and {Torre}(1991)]{bib:kuchar-torre-1991}
{Kucha\u{r}}, K.V.; {Torre}, C.G.
\newblock Gaussian reference fluid and interpretation of quantum
  geometrodynamics.
\newblock {\em Phys.\ Rev.\ D} {\bf 1991}, {\em 43},~419--441.
\newblock {\url{https://doi.org/10.1103/PhysRevD.43.419}}.

\bibitem[Mercuri and Montani(2004)]{bib:mercuri-montani-2004-framefixing}
Mercuri, S.; Montani, G.
\newblock {Revised} {Canonical} {Quantum} {Gravity} via the {Frame} {Fixing}.
\newblock {\em Int. J. Mod. Phys. D} {\bf 2004}, {\em
  13},~165--186.
\newblock {\url{https://doi.org/10.1142/s0218271804004359}}.

\bibitem[Montani et~al.(2011)Montani, Battisti, Benini, and
  Imponente]{bib:montani-primordialcosmology}
Montani, G.; Battisti, M.V.; Benini, R.; Imponente, G.
\newblock {\em Primordial Cosmology}; {World Scientific: Singapore}  2011.
\newblock {\url{https://doi.org/10.1142/7235}}.

\bibitem[Weinberg(2008)]{bib:weinberg}
Weinberg, S.
\newblock {\em Cosmology}; {OUP Oxford: Oxford, England}  2008.

\bibitem[Kolb and Turner(1990)]{bib:kolb-turner}
Kolb, E.W.; Turner, M.S.
\newblock {\em {The Early Universe}};  {Westview Press: Boulder, Colorado}  1990; Volume~69.
\newblock {\url{https://doi.org/10.1201/9780429492860}}.

\bibitem[Riotto(2002)]{bib:riotto-2017}
Riotto, A.
\newblock Inflation and the Theory of Cosmological Perturbations. \emph{arXiv}  \textbf{2002}, arXiv:hep-ph/0210162.
\newblock {\url{https://doi.org/10.48550/ARXIV.HEP-PH/0210162}}.

\bibitem[Arnowitt et~al.(1960)Arnowitt, Deser, and
  Misner]{bib:arnowitt-deser-misner-1960}
Arnowitt, R.; Deser, S.; Misner, C.W.
\newblock Canonical Variables for General Relativity.
\newblock {\em Phys. Rev.} {\bf 1960}, {\em 117},~1595--1602.
\newblock {\url{https://doi.org/10.1103/PhysRev.117.1595}}.

\bibitem[Misner et~al.(2017)Misner, Thorne, Wheeler, and
  Kaiser]{bib:misner-gravitation}
Misner, C.; Thorne, K.; Wheeler, J.; Kaiser, D.
\newblock {\em Gravitation}; {Princeton University Press: Princeton, New Jersey}  2017.

\bibitem[{Landau} and {Lifshitz}(1981)]{bib:landau-quantumMechanics}
{Landau}, L.D.; {Lifshitz}, E.M.
\newblock {\em Quantum Mechanics: Non-Relativistic Theory}, 3rd ed.; {
  Course on Theoretical Physics};  {Pergamon Press: Oxford, England}  1981; Volume~3.

\bibitem[Gundlach(1993)]{bib:gundlach-1993}
Gundlach, C.
\newblock Cosmological quantum fluctuations: Gauge-invariance and Gaussian
  states.
\newblock {\em Class. Quantum Gravity} {\bf 1993}, {\em 10},~1103.
\newblock {\url{https://doi.org/10.1088/0264-9381/10/6/007}}.

\bibitem[Maniccia et~al.(2023)Maniccia, Montani, and
  Antonini]{bib:maniccia-montani-antonini-2023}
Maniccia, G.; Montani, G.; Antonini, S.
\newblock {QFT} in curved spacetime from quantum gravity: Proper {WKB}
  decomposition of the gravitational component. \emph{Phys. Rev. D} \textbf{2023}, \emph{107}, L061901.
\newblock {\url{https://doi.org/10.1103/physrevd.107.l061901}}.

\bibitem[Brandenberger(2004)]{bib:brandenberger-book-2004}
Brandenberger, R.H. Lectures on the Theory of Cosmological Perturbations.
\newblock In {\em The Early Universe and Observational Cosmology}; Springer: Berlin/Heidelberg,  Germany, 2004; pp. 127--167.
\newblock {\url{https://doi.org/10.1007/978-3-540-40918-2_5}}.

\bibitem[Peter and Uzan(2013)]{bib:peter-uzan-PC}
Peter, P.; Uzan, J.P.
\newblock {\em {Primordial Cosmology}}; Oxford Graduate Texts, Oxford
  University {Press: Oxford, England}  2013.

\bibitem[Mukhanov(1985)]{bib:mukhanov-1985}
Mukhanov, V.F.
\newblock {Gravitational Instability of the Universe Filled with a Scalar
  Field}.
\newblock {\em JETP Lett.} {\bf 1985}, {\em 41},~493--496.

\bibitem[Sasaki(1986)]{bib:sasaki-1986}
Sasaki, M.
\newblock {Large Scale Quantum Fluctuations in the Inflationary Universe}.
\newblock {\em Prog. Theor. Phys.} {\bf 1986}, {\em
  76},~1036--1046.
\newblock {\url{https://doi.org/10.1143/PTP.76.1036}}.

\bibitem[Mukhanov(1988)]{bib:mukhanov-1988}
Mukhanov, V.F.
\newblock {Quantum Theory of Gauge Invariant Cosmological Perturbations}.
\newblock {\em Sov. Phys. JETP} {\bf 1988}, {\em 67},~1297--1302.

\bibitem[de~Blas and Olmedo(2016)]{bib:martin-olmedo-2016-LQC}
de~Blas, D.M.; Olmedo, J.
\newblock Primordial power spectra for scalar perturbations in loop quantum
  cosmology.
\newblock {\em J. Cosmol. Astropart. Phys.} {\bf 2016}, {\em
  2016},~029.
\newblock {\url{https://doi.org/10.1088/1475-7516/2016/06/029}}.

\bibitem[Li et~al.(2020)Li, Olmedo, Singh, and Wang]{bib:olmedo-singh-2020}
Li, B.F.; Olmedo, J.; Singh, P.; Wang, A.
\newblock Primordial scalar power spectrum from the hybrid approach in loop
  cosmologies.
\newblock {\em Phys. Rev. D} {\bf 2020}, {\em 102},~126025.
\newblock {\url{https://doi.org/10.1103/PhysRevD.102.126025}}.

\bibitem[Gielen and Mickel(2022)]{bib:gielen-2022-spectrum-bounce}
Gielen, S.; Mickel, L.
\newblock Gauge-Invariant Perturbations at a Quantum Gravity Bounce.
\newblock {\em Universe} {\bf 2022}, {\em 9},~29.
\newblock {\url{https://doi.org/10.3390/universe9010029}}.

\bibitem[Kiefer and Vardanyan(2022)]{bib:kiefer-tatevik-2022}
Kiefer, C.; Vardanyan, T.
\newblock Power spectrum for perturbations in an inflationary model for a
  closed universe.
\newblock {\em Gen. Relativ. Gravit.} {\bf 2022}, {\em 54},~30.
\newblock {\url{https://doi.org/10.1007/s10714-022-02918-3}}.

\bibitem[Cheng et~al.(2022)Cheng, Lee, and Ng]{bib:cheng-2022}
Cheng, S.L.; Lee, D.S.; Ng, K.W.
\newblock Power spectrum of primordial perturbations during ultra-slow-roll
  inflation with back reaction effects.
\newblock {\em Phys. Lett. B} {\bf 2022}, {\em 827},~136956.
\newblock {\url{https://doi.org/10.1016/j.physletb.2022.136956}}.

\bibitem[Bortolotti and Montani(2022)]{bib:bortolotti-2022}
Bortolotti, N.; Montani, G.
\newblock Inflationary Quantum Spectrum of the Quasi-Isotropic Universe. \emph{arXiv}  \textbf{2022}, arXiv:gr-qc/2212.08640.
   \url{http://xxx.lanl.gov/abs/2212.08640}.

\bibitem[Martin et~al.(2012)Martin, Vennin, and
  Peter]{bib:martin-vennin-peter-2012}
Martin, J.; Vennin, V.; Peter, P.
\newblock Cosmological inflation and the quantum measurement problem.
\newblock {\em Phys. Rev. D} {\bf 2012}, {\em 86},~103524.
\newblock {\url{https://doi.org/10.1103/PhysRevD.86.103524}}.

\bibitem[Langlois(1994)]{bib:langlois-1994}
Langlois, D.
\newblock Hamiltonian formalism and gauge invariance for linear perturbations
  in inflation.
\newblock {\em Class. Quantum Gravity} {\bf 1994}, {\em 11},~389.
\newblock {\url{https://doi.org/10.1088/0264-9381/11/2/011}}.

\bibitem[Giesel et~al.(2020)Giesel, Herold, Li, and Singh]{bib:giesel-2020}
Giesel, K.; Herold, L.; Li, B.F.; Singh, P.
\newblock {Mukhanov-Sasaki} equation in a manifestly gauge-invariant linearized
  cosmological perturbation theory with dust reference fields.
\newblock {\em Phys.\ Rev.\ D} {\bf 2020}, {\em 102},~023524.
\newblock {\url{https://doi.org/10.1103/physrevd.102.023524}}.

\bibitem[Kamenshchik et~al.(2021)Kamenshchik, Tronconi, and
  Venturi]{bib:venturi-2021}
Kamenshchik, A.Y.; Tronconi, A.; Venturi, G.
\newblock {The Born–Oppenheimer approach to quantum cosmology}.
\newblock {\em Class. Quantum Gravity} {\bf 2021}, {\em 38},~155011.
\newblock {\url{https://doi.org/10.1088/1361-6382/ac0a88}}.

\bibitem[Lewis(1967)]{bib:lewis-1967}
Lewis, H.R.
\newblock Classical and Quantum Systems with Time-Dependent
  Harmonic-Oscillator-Type Hamiltonians.
\newblock {\em Phys. Rev. Lett.} {\bf 1967}, {\em 18},~510--512.
\newblock {\url{https://doi.org/10.1103/PhysRevLett.18.510}}.

\bibitem[Lewis and Riesenfeld(1968)]{bib:lewis-1968}
Lewis, H.R.; Riesenfeld, W.B.
\newblock Class of Exact Invariants for Classical and Quantum Time‐Dependent
  Harmonic Oscillators.
\newblock {\em J. Math. Phys.} {\bf 1968}, {\em
  9},~1976--1986.
\newblock {\url{https://doi.org/10.1063/1.1664532}}.

\bibitem[Lewis and Riesenfeld(1969)]{bib:lewis-riesenfeld-1969}
Lewis, H.R.; Riesenfeld, W.B.
\newblock An Exact Quantum Theory of the Time‐Dependent Harmonic Oscillator
  and of a Charged Particle in a Time‐Dependent Electromagnetic Field.
\newblock {\em J. Math. Phys.} {\bf 1969}, {\em
  10},~1458--1473.
\newblock {\url{https://doi.org/10.1063/1.1664991}}.

\bibitem[Pedrosa(1997)]{bib:pedrosa-1997}
Pedrosa, I.A.
\newblock Exact wave functions of a harmonic oscillator with time-dependent
  mass and frequency.
\newblock {\em Phys. Rev. A} {\bf 1997}, {\em 55},~3219--3221.
\newblock {\url{https://doi.org/10.1103/PhysRevA.55.3219}}.

\bibitem[Bennett et~al.(2013)Bennett, Larson, Weiland, Jarosik, Hinshaw,
  Odegard, Smith, Hill, Gold, Halpern, Komatsu, Nolta, Page, Spergel, Wollack,
  Dunkley, Kogut, Limon, Meyer, Tucker, and Wright]{bib:wmap-2013}
Bennett, C.L.; Larson, D.; Weiland, J.L.; Jarosik, N.; Hinshaw, G.; Odegard,
  N.; Smith, K.M.; Hill, R.S.; Gold, B.; Halpern, M.;  et~al.
\newblock Nine-Year Wilkinson Microwave Anisotropy Probe (Wmap) Observations:
  Final Maps Additionally, Results.
\newblock {\em  Astrophys. J. Suppl. Ser.} {\bf 2013}, {\em
  208},~20.
\newblock {\url{https://doi.org/10.1088/0067-0049/208/2/20}}.

\bibitem[Ade et~al.(2016)]{bib:planck-results-2016}
Ade, P.A.R.  et~al. [Planck Collaboration]
\newblock PLANCK 2015 results.
\newblock {\em Astron. Astrophys.} {\bf 2016}, {\em 594},~A20.
\newblock {\url{https://doi.org/10.1051/0004-6361/201525898}}.

\bibitem[Aghanim et~al.(2020)]{bib:planck-results-2018}
Aghanim, N.  et~al. [Planck Collaboration]
\newblock {PLANCK 2018 results. V. CMB power spectra and likelihoods}.
\newblock {\em Astron. Astrophys.} {\bf 2020}, {\em 641},~A5.
\newblock {\url{https://doi.org/10.1051/0004-6361/201936386}}.

\bibitem[Cabass et~al.(2016)Cabass, Di~Valentino, Melchiorri, Pajer, and
  Silk]{bib:cabass-divalentino-melchiorri-2016}
Cabass, G.; Di~Valentino, E.; Melchiorri, A.; Pajer, E.; Silk, J.
\newblock Constraints on the running of the running of the scalar tilt from CMB
  anisotropies and spectral distortions.
\newblock {\em Phys. Rev. D} {\bf 2016}, {\em 94},~023523.
\newblock {\url{https://doi.org/10.1103/PhysRevD.94.023523}}.

\bibitem[van~de Bruck and Longden(2016)]{bib:carsten-2016}
van~de Bruck, C.; Longden, C.
\newblock Running of the running and entropy perturbations during inflation.
\newblock {\em Phys. Rev. D} {\bf 2016}, {\em 94},~021301.
\newblock {\url{https://doi.org/10.1103/PhysRevD.94.021301}}.

\bibitem[Nilsson(2022)]{bib:nilsson-2022}
Nilsson, N.A.
\newblock Explicit spacetime-symmetry breaking and the dynamics of primordial
  fields.
\newblock {\em Phys. Rev. D} {\bf 2022}, {\em 106},~104036.
\newblock {\url{https://doi.org/10.1103/PhysRevD.106.104036}}.

\bibitem[{Born} and {Oppenheimer}(1927)]{bib:born-1927}
{Born}, M.; {Oppenheimer}, R.
\newblock {Zur Quantentheorie der Molekeln}.
\newblock {\em Ann.  Phys.} {\bf 1927}, {\em 389},~457--484.
\newblock {\url{https://doi.org/10.1002/andp.19273892002}}.

\bibitem[Massar and Parentani(1998)]{bib:massar-1998}
Massar, S.; Parentani, R.
\newblock Particle creation and non-adiabatic transitions in quantum cosmology.
\newblock {\em Nucl. Phys. B} {\bf 1998}, {\em 513},~375--401.
\newblock {\url{https://doi.org/10.1016/S0550-3213(97)00718-9}}.

\bibitem[Schander and Thiemann(2021)]{bib:schander-thiemann-2021}
Schander, S.; Thiemann, T.
\newblock Backreaction in Cosmology.
\newblock {\em Front. Astron. Space Sci.} {\bf 2021}, {\em 8},~692198.
\newblock {\url{https://doi.org/10.3389/fspas.2021.692198}}.

\bibitem[Kiefer(2012)]{bib:kiefer-QG}
Kiefer, C.
\newblock {\em {Quantum Gravity}}, 3rd  ed.; Oxford University Press: New
  York,  NY, USA, 2012.

\bibitem[Mead(1992)]{bib:mead-1992-geomphase}
Mead, C.A.
\newblock The geometric phase in molecular systems.
\newblock {\em Rev. Mod. Phys.} {\bf 1992}, {\em 64},~51--85.
\newblock {\url{https://doi.org/10.1103/RevModPhys.64.51}}.

\bibitem[Panati et~al.(2007)Panati, Spohn, and Teufel]{bib:panati-2007}
Panati, G.; Spohn, H.; Teufel, S.
\newblock {The time-dependent Born-Oppenheimer approximation}.
\newblock {\em ESAIM M2AN} {\bf 2007}, {\em 41},~297--314.
\newblock {\url{https://doi.org/10.1051/m2an:2007023}}.

\bibitem[Min et~al.(2014)Min, Abedi, Kim, and Gross]{bib:seung-PRL-2014}
Min, S.K.; Abedi, A.; Kim, K.S.; Gross, E.K.U.
\newblock Is the Molecular Berry Phase an Artifact of the Born-Oppenheimer
  Approximation?
\newblock {\em Phys. Rev. Lett.} {\bf 2014}, {\em 113},~263004.
\newblock {\url{https://doi.org/10.1103/PhysRevLett.113.263004}}.

\bibitem[Bransden and Joachain(2003)]{bib:bransden}
Bransden, B.; Joachain, C.
\newblock {\em Physics of Atoms and Molecules}; {Prentice Hall: Hoboken, New Jersey}  2003.

\bibitem[Hwang and Noh(1996)]{bib:hwang-noh-1996}
Hwang, J.C.; Noh, H.
\newblock Cosmological perturbations in generalized gravity theories.
\newblock {\em Phys. Rev. D} {\bf 1996}, {\em 54},~1460--1473.
\newblock {\url{https://doi.org/10.1103/PhysRevD.54.1460}}.

\bibitem[Hwang(1997)]{bib:hwang-1997}
Hwang, J.C.
\newblock Cosmological perturbations in generalized gravity theories: Conformal
  transformation.
\newblock {\em Class. Quantum Gravity} {\bf 1997}, {\em 14},~1981.
\newblock {\url{https://doi.org/10.1088/0264-9381/14/7/029}}.

\bibitem[Hwang and Noh(2001)]{bib:hwang-noh-2001}
Hwang, J.C.; Noh, H.
\newblock Gauge-ready formulation of the cosmological kinetic theory in
  generalized gravity theories.
\newblock {\em Phys. Rev. D} {\bf 2001}, {\em 65},~023512.
\newblock {\url{https://doi.org/10.1103/PhysRevD.65.023512}}.

\bibitem[Hwang and Noh(2005)]{bib:hwang-noh-2005}
Hwang, J.C.; Noh, H.
\newblock Classical evolution and quantum generation in generalized gravity
  theories including string corrections and tachyons: Unified analyses.
\newblock {\em Phys. Rev. D} {\bf 2005}, {\em 71},~063536.
\newblock {\url{https://doi.org/10.1103/PhysRevD.71.063536}}.

\bibitem[Capozziello et~al.(2008)Capozziello, {De Laurentis}, and
  Francaviglia]{bib:capozziello-2008}
Capozziello, S.; {De Laurentis}, M.; Francaviglia, M.
\newblock Higher-order gravity and the cosmological background of gravitational
  waves.
\newblock {\em Astropart. Phys.} {\bf 2008}, {\em 29},~125--129.
\newblock {\url{https://doi.org/10.1016/j.astropartphys.2007.12.001}}.

\bibitem[Capozziello et~al.(2013)Capozziello, Carlevaro, Laurentis, Lattanzi,
  and Montani]{bib:capozziello-montani-2011}
Capozziello, S.; Carlevaro, N.; Laurentis, M.; Lattanzi, M.; Montani, G.
\newblock {Cosmological implications of a viable non-analytical f(R) model}.
\newblock {\em Eur. Phys. J. Plus} {\bf 2013}, {\em 128},~155,
  \href{http://xxx.lanl.gov/abs/1104.2169}{{\normalfont
  [arXiv:astro-ph.CO/1104.2169]}}.
\newblock {\url{https://doi.org/10.1140/epjp/i2013-13155-4}}.

\bibitem[Bamonti et~al.(2022)Bamonti, Costantini, and
  Montani]{bib:bamonti-costantini-montani-2022}
Bamonti, N.; Costantini, A.; Montani, G.
\newblock Features of the primordial Universe in f(R)-gravity as viewed in the
  Jordan frame.
\newblock {\em Class. Quantum Gravity} {\bf 2022}, {\em 39},~175011.
\newblock {\url{https://doi.org/10.1088/1361-6382/ac7694}}.

\bibitem[De~Angelis et~al.(2021)De~Angelis, Figurato, and
  Montani]{bib:deangelis-figurato-montani-2021}
De~Angelis, M.; Figurato, L.; Montani, G.
\newblock Quantum dynamics of the isotropic universe in metric $f(R)$ gravity.
\newblock {\em Phys. Rev. D} {\bf 2021}, {\em 104},~024054.
\newblock {\url{https://doi.org/10.1103/PhysRevD.104.024054}}.

\bibitem[Angelis and Montani(2022)]{bib:montani-deangelis-2022}
Angelis, M.D.; Montani, G.
\newblock On the emergence of a classical Isotropic Universe from a Quantum
  $f(R)$ Bianchi Cosmology in the Jordan Frame. \emph{arXiv}  \textbf{2022}, arXiv:gr-qc/2207.14683.
  \url{http://xxx.lanl.gov/abs/2207.14683}.

\end{thebibliography}
\end{document}